\documentclass[a4paper,11pt]{article}
\usepackage{jheppub}
\usepackage[T1]{fontenc}
\usepackage{graphicx}
\usepackage{amsmath}
\usepackage{amsmath}
\usepackage{xspace}
\usepackage{subfig}
\usepackage{color}
\usepackage[utf8]{inputenc}
\usepackage{commath}
 \usepackage{cancel}
\usepackage{slashed}
\usepackage{fancyvrb}
\usepackage{multirow}

 \notoc
\title{\boldmath Heavy Neutrinos at the FCC-hh\\ in the $U(1)_{B-L}$ Model}

\author[a,b]{Wei Liu,}
\author[c]{Suchita Kulkarni,}
\author[b]{Frank F. Deppisch.}

\affiliation[a]{Department of Applied Physics, Nanjing University of Science and Technology, Nanjing 210094, China}
\affiliation[b]{University College London, Gower Street, London WC1E 6BT, UK}
\affiliation[c]{Institute of Physics, NAWI Graz, University of Graz,Universit\"atsplatz 5, A-8010 Graz, Austria}

\emailAdd{wei.liu@njust.edu.cn, wei.liu.16@ucl.ac.uk}
\emailAdd{f.deppisch@ucl.ac.uk}
\emailAdd{suchita.kulkarni@uni-graz.at}

\abstract{We investigate the potential of the 100 TeV future circular collider (FCC-hh) to probe heavy neutrinos. We concentrate in particular on heavy neutrino production via a $U(1)_{B-L}$ $Z'$ gauge boson and contrast the resulting limits with that mediated by Standard Model weak currents. We consider heavy neutrino decays to semi-leptonic as well as fully leptonic final states, particularly with muon flavour, and we show the importance of considering searches both in prompt and displaced decays of the heavy neutrinos. For prompt final states, semi-leptonic modes are more promising due to smaller background and larger yields, and TeV-scale heavy neutrinos with active-sterile mixing compatible with light neutrino mass generation in a seesaw scenario can be probed for a 5 TeV $Z'$ and gauge coupling as low as $g_{B-L} = 10^{-2}$. Displaced vertex searches can extend this range to heavy neutrino masses as low as 10 GeV.}

\begin{document}
\maketitle
\flushbottom

%%%%%%%%%%%%%%%%%%%%%%%%%%%%%%%%%%%%%%%%%%%%%%%%%%%%%%%%%%%%%%%%%%%%%%%%%%%%%%%%%
\section{Introduction}
%%%%%%%%%%%%%%%%%%%%%%%%%%%%%%%%%%%%%%%%%%%%%%%%%%%%%%%%%%%%%%%%%%%%%%%%%%%%%%%%%
The observation of small neutrino masses, which have no explanation within the Standard Model (SM) of particle physics, points towards new physics. Among possible scenarios which can explain neutrino masses, the $U(1)_{B-L}$ gauge model is one of the simplest anomaly free constructions~\cite{Davidson:1978pm,Mohapatra:1980qe}. It incorporates three right-handed (RH) neutrinos $N_i$ which acquire Majorana masses upon the spontaneous breaking of the $U(1)_{B-L}$ gauge symmetry. In turn, this induces light active neutrino masses via the seesaw mechanism. As opposed to the minimal seesaw scenario, the RH neutrinos are not completely sterile under the model's gauge symmetry. Instead, they are charged under $U(1)_{B-L}$ and they couple to the associated heavy gauge boson $Z'$ and $B-L$ charged Higgs $\chi$ present in the model. This opens up a rich phenomenology as it not only provides the SM charged and neutral current portals for heavy neutrino production but also production via the $Z'$ and $\chi$.

The $B-L$ production mechanisms are not suppressed by the active-sterile mixing which is generically expected to be very small, $|V_{l N}| \sim \sqrt{m_\nu / m_N} \lesssim 10^{-6}\times(100~\text{GeV} / m_N)$, to accommodate the observed light neutrino masses $m_\nu \lesssim 0.1$~eV. For such a small active-sterile mixing, the heavy neutrinos, which can only decay through such suppressed channels, are long-lived with a proper decay length $L_N^0 \sim 2.5~\text{cm}\times(10^{-6}/|V_{l N}|)^2(100~\text{GeV} / m_N)^5$ for $m_N \lesssim 100$~GeV. This naturally leads to displaced vertex signatures at colliders for heavy neutrino masses $m_N \approx 100$~GeV.

Searches for RH neutrinos via SM mediated processes are being carried out at the LHC, including both prompt and displaced final states~\cite{Chatrchyan:2012fla, Aaij:2014aba, Aad:2015xaa, CMS:2015qur, CMS:2016aro, CortinaGil:2017mqf, Mermod:2017ceo, Izmaylov:2017lkv, Sirunyan:2018mtv}. These searches currently put an upper limit of $|V_{\mu N}| \lesssim 10^{-3}$ on the active-sterile mixing with the muon and RH neutrino masses $m_N \lesssim$~100 GeV. The limits weaken rapidly for increasing heavy neutrino masses. This is expected because the SM $W$ becomes increasingly off-shell and thus the production cross section is doubly suppressed, from the active-sterile mixing as well as the mass of the heavy RH neutrino. Further projections for RH neutrinos from the SM production, at the LHC~\cite{Helo:2010cw, Liventsev:2013zz, Abada:2013aba, Helo:2013esa, Canetti:2014dka, Gago:2015vma, Das:2015toa, Banerjee:2015gca, Izaguirre:2015pga, Arganda:2015ija, Antusch:2015mia, Degrande:2016aje, Antusch:2017pkq, Ruiz:2017yyf, Antusch:2017hhu, Dube:2017jgo, Cai:2017mow, Deppisch:2018eth, Abada:2018sfh, Cottin:2018nms, Drewes:2018gkc, Dib:2018iyr, Boiarska:2019jcw, Cheung:2020buy, Jones-Perez:2019plk, Liu:2019ayx, Drewes:2019fou}, the proposed SHiP detector~\cite{Alekhin:2015byh, SHiP:2018xqw}, LHeC~\cite{Das:2018usr} and the FCC~\cite{Antusch:2016ejd, Pascoli:2018heg, Antusch:2015mia, Antusch:2017pkq} have been carried out. These studies rely on the minimal and ensured RH neutrino production mechanism, taking advantage of higher luminosity, larger detectors and increased center-of-mass energies. They remain fundamentally limited by the right-handed neutrino production cross-section suppressed by the active-sterile mixing. Together with the stringent cuts on the final state objects at FCC-hh, the studies show that probing the seesaw neutrino mass generation mechanism at future colliders via the minimal production channels will be very challenging.

This motivates the consideration of non-minimal models and probing neutrino mass generation mechanisms via exotic production modes. In this context, the $B-L$ model offers an interesting avenue as it contains three additional heavy neutrino generation portals unsuppressed by the active-sterile mixing namely the heavy $Z'$, heavy $B-L$ Higgs and the SM Higgs. There are two main advantages of these additional modes, especially for $Z'$ mediated processes. Firstly, the right-handed neutrinos can have large $p_T$ depending on the $Z'$ mass which can be produced on-shell. Secondly, given that this on-shell $Z'$ can be heavier than the $W$ boson, right-handed neutrino masses $m_N \gtrsim 100 \,\rm{GeV}$  can be probed without penalising production cross sections as opposed to the SM case where this mass range necessitates off-shell $W$ mediator and leads to an additional suppression in cross section. These considerations enable searching for right-handed neutrinos in regions of $|V_{\mu N}|$ and $m_N$ parameter space otherwise inaccessible. Analyses of $B-L$ processes have been performed in the literature. This includes reinterpretations of existing $Z'$ production searches~\cite{Chiang:2019ajm}, analyses of the sensitivity at the lifetime frontier~\cite{Deppisch:2013cya, Batell:2016zod, Deppisch:2019kvs, Bhattacherjee:2021rml}, explorations at the LHC~\cite{Accomando:2017qcs, Das:2019fee, Cheung:2021utb, FileviezPerez:2020cgn, Accomando:2016rpc},  as well as prompt searches at the FCC-hh~\cite{Han:2021pun}.

In this work, we look ahead and consider the potential of the FCC-hh detector at 100~TeV center-of-mass energy to probe right-handed neutrino production via the $B-L$ $Z'$. We contrast this mode with processes mediated by SM weak currents. In doing so, we not only exploit the gain in the center-of-mass energy and the higher luminosity (30~ab$^{-1}$), but also the larger detector volume capable of capturing longer lived right-handed neutrinos. In doing such an exercise our aim is to demonstrate the complementarity between SM weak current and $B-L$ mediated processes as well as to understand the prominent kinematic differences between FCC-hh and LHC, the two SM and $B-L$ channels. We consequently illustrate regions of parameter space which can be probed by simple analyses while highlighting the necessity to develop more comprehensive analyses techniques.

The paper is organised as follows: We begin by reviewing the $B-L$ gauge model and its immediate phenomenological consequences in Section~\ref{blreview}. This is followed by a discussion of our analysis setup in Section~\ref{sec:analysis_setup}. Using this as a basis, in Section~\ref{sec:zpreach} we first evaluate the sensitivity of FCC-hh for $Z'$ resonance production. We then discuss heavy neutrino production via SM $W$ and $B-L$ $Z'$ processes in Section~\ref{sec:HPD}. We show our sensitivity estimates in Section~\ref{sec:sensitivity} and finally conclude in Section~\ref{sec:conclusion}.

%%%%%%%%%%%%%%%%%%%%%%%%%%%%%%%%%%%%%%%%%%%%%%%%%%%%%%%%%%%%%%%%%%%%%%%%%%%%%%%%%
\section{The $B-L$ gauge model}
\label{blreview}
%%%%%%%%%%%%%%%%%%%%%%%%%%%%%%%%%%%%%%%%%%%%%%%%%%%%%%%%%%%%%%%%%%%%%%%%%%%%%%%%%
\subsection{Model setup and particle spectrum}
%%%%%%%%%%%%%%%%%%%%%%%%%%%%%%%%%%%%%%%%%%%%%%%%%%%%%%%%%%%%%%%%%%%%%%%%%%%%%%%%%
In addition to the particle content of the SM, the $U(1)_{B-L}$ model contains an Abelian gauge field $B^\prime_\mu$, a SM singlet scalar field $\chi$ and three RH neutrinos $\nu_{R,i}$. The gauge group is $SU(3)_C\times SU(2)_L \times U(1)_Y \times U(1)_{B-L}$, where $\chi$ and $N_i$ have $B-L$ charges $B-L = +2$ and $-1$, respectively. All SM particles carry their conventional $B-L$ quantum number. The scalar potential includes all terms allowed by the symmetry,
\begin{align}
\label{VHX}
	{\cal V}(H,\chi) = m^2 H^\dagger H + \mu^2 |\chi|^2 + \lambda_1 (H^\dagger H)^2
	          + \lambda_2 |\chi|^4 + \lambda_3 H^\dagger H |\chi|^2,
\end{align}
with the SM Higgs doublet $H$. The scalar sector consists of a light SM-like Higgs $h_1 \sim h_0^\text{SM}$ with $m_{h_1} \approx 125$~GeV and a heavy Higgs $h_2$ with $m_{h_2} \approx \sqrt{2\lambda_2}\langle\chi\rangle$. Here, $\langle\chi\rangle$ is the vacuum expectation value (VEV) of the $B-L$ Higgs $\chi$. The mixing between the Higgs fields, induced by the term $\lambda_3$ in Eq.~\eqref{VHX}, can still be sizeable but is not relevant for our discussion. We denote the gauge coupling strength associated with the $U(1)_{B-L}$ symmetry by $g_{B-L}$. In this paper, we neglect a kinetic mixing between the $U(1)_{B-L}$ and $U(1)_Y$ gauge bosons, i.e. we consider the minimal $B-L$ gauge model. Such a mixing will be generated radiatively even if assumed zero at, e.g., the $B-L$ breaking scale, $\epsilon(m_Z) \sim \frac{e g_{B-L}}{16\pi^2}\log(m_{Z'}^2/m_Z^2)$. This is negligible for our analysis. We note that a finite kinetic mixing can be beneficial in searching for heavy neutrinos. Its main effect will be that the SM $Z$ can decay to two heavy neutrinos, suppressed by $\epsilon$, but not by the active-sterile neutrino mixing. The mass of the $Z'$ gauge boson is then simply given by $m_{Z'} = g_{B-L} \langle\chi\rangle$.

The model contains the additional Yukawa terms
\begin{align}
\label{LY}
	{\cal L} \supset
	   - y_{ij}^\nu \overline{L_i}\nu_{R,j}\tilde{H}
	   - y_{ij}^M \overline{\nu^c_{R,i}} \nu_{R,j}\chi
	   + \text{h.c.},
\end{align}
where $L_i$ are the SM lepton doublets, $\tilde{H} = i\sigma^2 H^*$ and a summation over the generation indices $i,j = 1, 2, 3$ is implied. The Yukawa matrices $y^\nu$ and $y^M$ are a priori arbitrary. The RH neutrino masses are generated by breaking of the $B-L$ symmetry, with the mass matrix given by $M_R = \sqrt{2}y^M\langle\chi\rangle$. The light neutrinos mix with the RH neutrinos via the Dirac mass matrix $m_D = y^\nu v/\sqrt{2}$ where $v = \langle H^0 \rangle \approx 246$~GeV is the SM Higgs VEV. The combined $6\times 6$ mass matrix in the $(\nu_L, \nu_R^c)$ basis is then in block form
\begin{align}
\label{MD}
	{\cal M} =
	\begin{pmatrix}
		0   & m_D \\
		m_D & M_R.
\end{pmatrix},
\end{align}
In the seesaw limit we consider, $||M_R|| \gg ||m_D||$, the light and heavy neutrino mass matrices are $m_\nu \sim - m_D \cdot M^{-1}_R \cdot m^T_D$ and $m_N \sim M_R$, respectively. The flavour $(\nu_{L,i}, \nu^c_{R,i})$ and mass $(\nu_i, N_i)$ eigenstates of the light and heavy neutrinos are connected in block form as
\begin{align}
\label{Neutrino}
	\begin{pmatrix}
		\nu_L \\ \nu^c_R
	\end{pmatrix} =
	\begin{pmatrix}
		V_{LL} & V_{LR} \\
		V_{RL} & V_{RR}
	\end{pmatrix} \cdot
	\begin{pmatrix}
		\nu \\ N
	\end{pmatrix}.
\end{align}
The mixing matrix $V_{LL}$ and the light neutrino masses are constrained by oscillation experiments to yield their observed values, i.e. the SM charged current lepton mixing $V_{LL} \approx U_\text{PMNS}$ (apart from small non-unitarity corrections and taking the basis in which the charged lepton mass matrix is diagonal). The approximately unitary matrix $V_{RR}$ describes the mixing among the RH neutrinos and the active-sterile mixing is $V_{LR} \approx m_D \cdot M_R^{-1}$.

We effectively consider the case of a single RH neutrino generation mixing with one SM neutrino at a time; specifically we focus on the mixing with the muon neutrino and hence take $V_{LL}, V_{RR} \sim 1$ and $V_{LR} \sim V_{\mu N}$ with active-sterile mixing strength $V_{\mu N}$ suppressing the charged-current interaction of the muon with the RH neutrino. We take the RH neutrino mass $m_N$ and the mixing strength $V_{\mu N}$ as free model parameters. In order to generate a light neutrino mass $m_\nu \lesssim 0.1$~eV via the seesaw mechanism with $||m_D|| \ll ||M_R||$, the mixing strength takes the generic value
\begin{align}
	|V_{\mu N}| \approx \frac{m_D}{M_R} = \sqrt{\frac{m_\nu}{m_N}}
	            = 10^{-6}\times \left(\frac{100\,\text{GeV}}{m_N}\right).
\end{align}
This is the generic expectation for the mixing strength of a single heavy Majorana neutrino generating a light neutrino mass of 0.1~eV. Both smaller and larger mixing is possible in more realistic and extended scenarios that incorporate all three generations of light neutrinos and at least two heavy neutrinos. Smaller mixing is possible by decoupling the heavy neutrino with other heavy neutrinos responsible for light neutrino mass generation. Larger mixing, up to current experimental constraints, is possible if there is cancellation among contributions of different heavy neutrinos. This can be achieved with either three generations of heavy neutrinos (see, e.g., \cite{Deppisch:2010fr}) or extended scenarios with quasi-Dirac heavy neutrinos, usually referred to as inverse seesaw.

%%%%%%%%%%%%%%%%%%%%%%%%%%%%%%%%%%%%%%%%%%%%%%%%%%%%%%%%%%%%%%%%%%%%%%%%%%%%%%%%%
\subsection{Production and decay of heavy neutrinos}
\label{sec:prodN}
%%%%%%%%%%%%%%%%%%%%%%%%%%%%%%%%%%%%%%%%%%%%%%%%%%%%%%%%%%%%%%%%%%%%%%%%%%%%%%%%%
Within the $B-L$ model, heavy neutrinos can be produced at the FCC-hh via different mechanisms. The first and foremost are the SM mediators, i.e. the $W$ and $Z$ gauge bosons and the Higgs $h_1$. In addition, there are the $B-L$ mediators, namely the $Z'$ gauge boson and the heavy Higgs boson $h_2$.

Out of these mechanisms, the SM Higgs production $pp \to h_1 \to N N$ will be suppressed by the Higgs mixing and the mass of the heavy neutrino with respect to the $B-L$ breaking scale, $m_N / \langle\chi\rangle$ ($m_N < m_{h_1}/2$) \cite{Deppisch:2018eth}. The alternative mode $pp \to h_1 \to N\nu$ is instead suppressed by the active-sterile neutrino mixing. The production via the heavy Higgs $h_2$, $pp \to h_2 \to NN$, is possible as well but likewise suppressed by $m_N / \langle\chi\rangle$, though higher neutrino masses can be probed. The mode $pp \to h_2 \to N\nu$ is suppressed by both the Higgs and the active-sterile neutrino mixing.

The best studied SM modes are via the $W$ and $Z$ gauge bosons, $pp\to W\to N\ell$ and $pp\to Z\to N\nu$. They are generally suppressed by the active-sterile neutrino mixing and for $m_N > m_W, m_Z$ proceed through off-shell gauge bosons with further strong suppression for high $N$ masses. The $W$ mediated channel gives rise to a prompt lepton which helps in accepting events even if the heavy neutrino decays via a displaced vertex. We will thus focus on this mode.

Finally, the main production mode in our analysis is via the $B-L$ $Z'$ gauge boson, $pp \to Z' \to NN$. The production cross section $\sigma(pp\to Z')$ at the LHC and FCC-hh is shown in Fig.~\ref{fig:zp_limits}~(left) in Section~\ref{sec:zpreach} to determine the reasonable range of the gauge coupling $g_{B-L}$ at the FCC-hh era. The partial decay width of $Z'\to NN$ per generation of heavy neutrino is
\begin{align}
\label{partialdecay}
	\Gamma(Z'\to NN)
	= g_{B-L}^2\frac{m_{Z'}}{24\pi} \left(1 - \frac{4 m_N^2}{m_{Z'}^2}\right)^{3/2}.
\end{align}
Considering the $Z'$ to be much heavier than the SM fermions into which it decays, the total decay width is
\begin{align}
\label{totaldecay}
	\Gamma(Z') \approx g_{B-L}^2\frac{m_{Z'}}{24\pi}
	\left[13 + 3\left(1-\frac{4m_N^2}{m_{Z'}^2}\right)^{3/2}\right],
\end{align}
with three generations of heavy neutrinos.

In all mechanisms we must then consider the decays of the heavy neutrino. They proceed via the same mediators as discussed above. The $h_1, h_2$ mediated decays are suppressed by the Yukawa coupling $\sim m_N/\langle\chi\rangle$ and the heavy neutrino dominantly decays via SM $W$ and $Z$, $N\to W^{(*)}\ell, Z^{(*)}\nu$. We do not consider decays via $Z'$ such as $N_2\to Z' N_1$ as we take into account only a single generation of heavy neutrino. In any case the gauge boson mediated decay widths are suppressed by the active-sterile mixing.

For heavy neutrino masses $m_N \lesssim m_W, m_Z$, decays will take place via off-shell $W$ and $Z$. The branching ratios of such light heavy neutrinos are discussed in \cite{Atre:2009rg, Deppisch:2018eth}. If the heavy neutrino mass is above this threshold, two body decays via on-shell $W$, $Z$ occur. The heavy neutrino branching ratios in such cases are approximately independent of the mass and $BR(N\to W\ell)\approx 70\%$ while $BR(N\to Z\nu)\approx 30\%$. Considering decays through a single SM lepton generation, the branching ratios are also independent of the active-sterile mixing.

With the SM $W$ and $Z$ decaying further, the $N$ decays can thus result in $N\to\ell^+ \ell^- \nu$ corresponding to the $W^{(*)}$ and $Z^{(*)}$ mediated leptonic decays with visible final states. Finally, we also get $N\to\ell j j $ which corresponds to the hadronic $W$, $Z$ decays.

With the assumption of flavour-diagonal active-sterile neutrino mixing, we concentrate on one generation of heavy neutrinos, namely coupling to muons and muon neutrinos. Out of the several final states it can produce we will in particular consider $N\to \mu\mu\nu$, $N\to \mu jj$ for neutrino masses lighter than 100~GeV and $N\to \mu W \to \mu \mu\nu$, $N\to \mu W \to\mu jj$ for heavier neutrinos. We concentrate on  muon flavour as muon performance in several parts of the detector is generally expected to be superior and we consider the $\mu jj$ final state as it has a larger branching ratio.

While the branching ratios are largely independent of the heavy neutrino mass and the active-sterile neutrino mixing, the total decay width and thus the decay length crucially depend on it. For $m_N\lesssim m_Z$, the proper decay length can be approximated as \cite{Atre:2009rg}
\begin{align}
\label{lengthapproxi}
	L^0_N \approx 2.5~\text{cm}
	\times\left(\frac{10^{-6}}{|V_{\mu N}|}\right)^2
	\times\left(\frac{100~\text{GeV}}{m_N}\right)^5.
\end{align}
For $m_N \gtrsim 100$~GeV, decays are proportionally faster due to their on-shell nature. Approximately, the proper decay length in this regime is \cite{Atre:2009rg}
\begin{align}
	\label{lengthapproxi-onshell}
	L^0_N \approx 0.1~\text{mm}
	\times\left(\frac{10^{-6}}{|V_{\mu N}|}\right)^2
	\times\left(\frac{100~\text{GeV}}{m_N}\right)^3.
\end{align}
%

%%%%%%%%%%%%%%%%%%%%%%%%%%%%%%%%%%%%%%%%%%%%%%%%%%%%%%%%%%%%%%%%%%%%%%%%%%%%%%%%%
\section{Analysis setup}
\label{sec:analysis_setup}
%%%%%%%%%%%%%%%%%%%%%%%%%%%%%%%%%%%%%%%%%%%%%%%%%%%%%%%%%%%%%%%%%%%%%%%%%%%%%%%%%
\subsection{FCC-hh detector geometry}
\label{sec:FCC}
%%%%%%%%%%%%%%%%%%%%%%%%%%%%%%%%%%%%%%%%%%%%%%%%%%%%%%%%%%%%%%%%%%%%%%%%%%%%%%%%%
The future physics collider program has two major goals; first and foremost to measure the SM Higgs boson and SM electroweak sector properties as precisely as possible and a second to search for and potentially discover new physics which may be out of reach at the LHC. A two stage plan is proposed towards fulfilling these goals. The first stage is an electron-positron collider with a center-of-mass collision energy of 240 GeV called the FCC-ee\footnote{In addition, there will also be a run at the $Z$-pole with luminosity larger than the LEP luminosity.}. The second will be a hadron-hadron collider with a center-of-mass collision energy of 100 TeV and an integrated luminosity of at least 10 times larger of the HL-LHC, i.e., 30 ab$^{-1}$~\cite{FCC:2018vvp}. In this paper, we focus on the reach of FCC-hh for a BSM particle, the heavy neutrino, therefore, we briefly discuss below the FCC-hh geometry.
\begin{figure}[t!]
\centering
\includegraphics[width=0.99\textwidth]{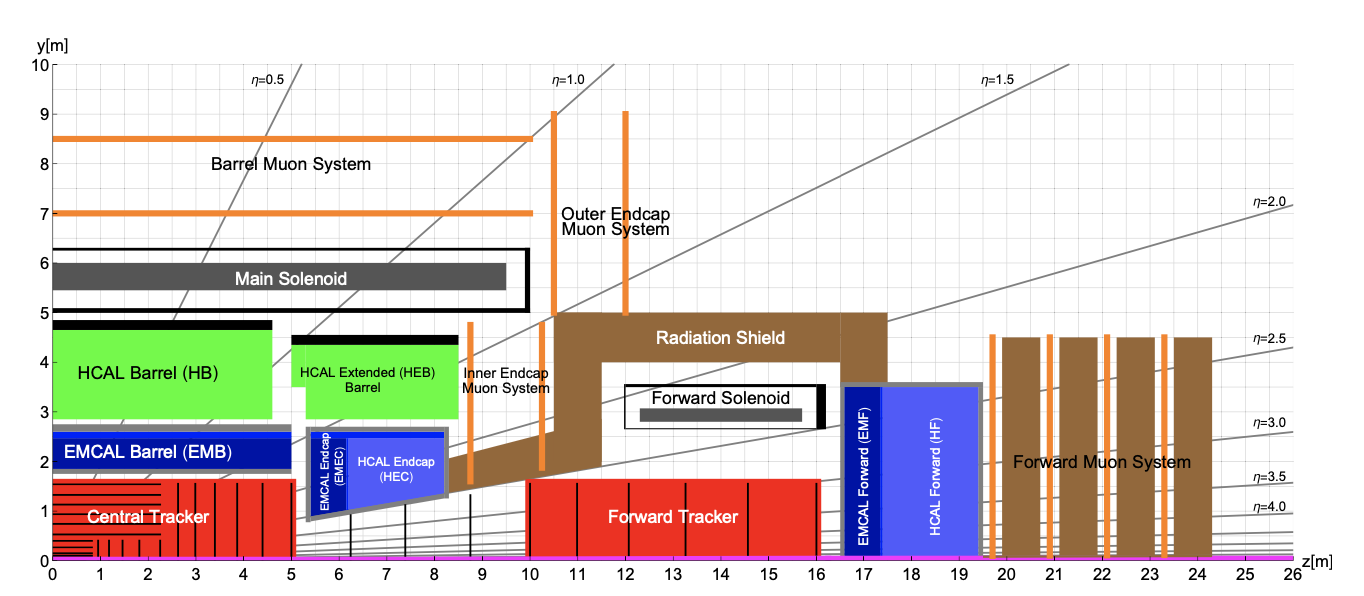}
\caption{Longitudinal cross-section of the FCC-hh reference detector. Taken from \cite{FCC:2018vvp}.}
\label{fig:detector}
\end{figure}

Due to the high center-of-mass energy, the FCC-hh can in general produce boosted objects, therefore the FCC-hh detector is being designed to accept high pseudo-rapidities $|\eta| \lesssim 4$. The detector subcomponents and geometry are shown in Fig.~\ref{fig:detector}, taken from Ref.~\cite{FCC:2018vvp}. As detailed below we implement this geometry in our analysis in order to account for geometrical acceptance. Compared to the LHC detector, the FCC-detector is larger both in longitudinal and transverse directions in order to capture higher energy final states. The larger volume is of a particular importance in detecting long lived particles as larger lifetimes can be probed.

In order to estimate the potential of the FCC-hh detector to probe displaced final states, we consider a fiducial volume of the detector, in particular the regions~\cite{FCC:2018vvp}:
\begin{itemize}
	\item Inner tracker: $0.025~\text{m} < L_{xy} < 1.55$~m and $L_z < 5$~m,
	\item Region 2 (calorimeter): $1.7~\text{m} < L_{xy} < 7$~m and $L_z < 9$~m,
	\item Forward tracker: $2.5<|\eta|<4$, $0.025~\text{m} < L_{xy} < 1.55$~m and $10~\text{m} < L_z < 16$~m,
	\item Forward Region 2 (calorimeter):\\
	$2.5<|\eta|<4$, $0.025~\text{m} < L_{xy} < 4$~m and $16.5~\text{m} < L_z < 19.5$~m.
\end{itemize}
here $L_{xy}$ and $L_{z}$ are the transverse and perpendicular displacements respectively. We consider a decay prompt if $L_N \leq 1$~mm~\cite{FCC:2018vvp}. This is possible either when the mixing  $|V_{\mu N}|$ is large or $m_N > 100$ GeV, irrespective of the SM or $B-L$ mediated production channels. 

%%%%%%%%%%%%%%%%%%%%%%%%%%%%%%%%%%%%%%%%%%%%%%%%%%%%%%%%%%%%%%%%%%%%%%%%%%%%%%%%%
\subsection{Signal final states}
\label{sec:sig_fin_states}
%%%%%%%%%%%%%%%%%%%%%%%%%%%%%%%%%%%%%%%%%%%%%%%%%%%%%%%%%%%%%%%%%%%%%%%%%%%%%%%%%

\begin{figure}[t!]
	\centering
	\includegraphics[width=0.9\textwidth]{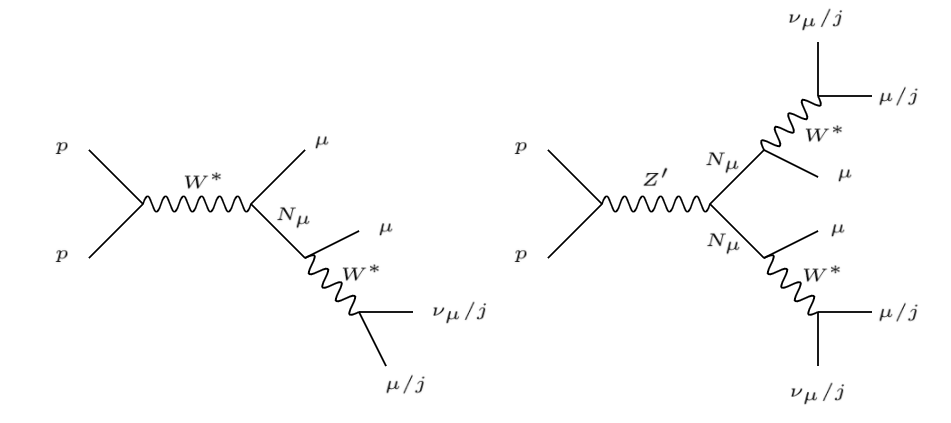}
	\caption{Feynman diagram of the SM $W$ production of the heavy neutrino, $pp \to W^* \to N_{\mu} \mu$~(left) and $Z'$ production of the heavy neutrino, $pp \to Z^{\prime} \to N N$~(right). }
	\label{fig:feynman}
\end{figure}

A number of different final states are possible depending on the production and decay of the heavy neutrino. We show the corresponding Feynman diagrams in Fig.~\ref{fig:feynman}, which includes the full production and decay chain for the RH neutrinos we consider in this work. The corresponding final states are summarised in Table ~\ref{tab:sig_category}. In general they contain a combination of muons, jets and missing energy. We separate the processes depending on the production mechanism i.e. SM or $B-L$, and also on the prompt or displaced category. Among the prompt production, five different final states are possible. Out of these final states, within this analysis we will not consider $3\mu + 2j + \cancel{E}_T$ final state.

We foresee a signal - background discrimination on the basis of missing energy where available. When missing energy is not available, we will demonstrate the use of same sign (SS) or opposite sign (OS) muons in the final state. Later on we will explicitly demonstrate the effectiveness of these strategies to derive the final sensitivity. For the displaced final states, we will consider an inclusive analysis considering either one prompt lepton and one displaced vertex originating out of two displaced muons (for SM mediated heavy neutrino production), or one displaced vertex forming out of two displaced muons (for $B-L$ mediated production). 

\begin{table}[b!]
\centering
 \begin{tabular}{| c |c | c |} 
 \hline
Category & Decay chain & Final state \\
 \hline
 SM  & $p p \to W^{*} \to \mu N; N \to \mu \mu \nu$ & 3$\mu + \cancel{E}_T$\\ 
 SM  & $p p \to W^{*} \to \mu N; N \to \mu j j$ & $2\mu + 2j$ (OS/SS)\\
 $B-L$  & $p p \to Z^{\prime} \to N N; N \to \mu j j; N \to \mu j j$ & $2\mu + 4j$ (OS/SS)\\ 
 $B-L$  & $p p \to Z^{\prime} \to N N; N \to \mu j j; N \to \mu \mu \nu$ & $3\mu + 2j + \cancel{E}_T$\\
 $B-L$  & $p p \to Z^{\prime} \to N N; N \to \mu \mu \nu; N \to \mu \mu \nu$ &  $4\mu + \cancel{E}_T$\\
 \hline
 \hline
 SM  & $p p \to W^* \to \mu N; N \to \mu \mu \nu$ & $\mu^{\rm{prompt}}$ + $2\mu^{\rm{disp}}$\\ 
 SM  & $p p \to W^* \to \mu N; N \to \mu j j$ & $\mu^{\rm{prompt}}$ + $(\mu j j)^{\rm{disp}}$\\ 
 $B-L$  & $p p \to Z \to N N; N \to \mu \mu \nu$; $N \to$ inclusive & $2\mu^{\rm{disp}}$\\
 $B-L$  & $p p \to Z \to N N; N \to \mu j j$; $N \to$ inclusive & $(\mu j j)^{\rm{disp}}$\\
 \hline
 \end{tabular}
\caption{Different final states arising from prompt decays of heavy neutrinos for SM and $B-L$ mediated production mechanism.} \label{tab:sig_category}
\end{table}
In order to trigger on this signal, we use either a single prompt or displaced lepton trigger.  In accordance with the FCC-hh Conceptual Design Report (CDR)~\cite{FCC:2018vvp}, we require the leading muon $p_T$ greater than 150 GeV mimicking a trigger strategy. For a displaced muon trigger, no concrete numbers are as yet available, however we increase the $p_T$ cut by a factor 1.5 in order to keep in line with the experience of dealing with displaced versus prompt objects at the LHC~\cite{private}. Thus, for a displaced muon final state, we require a leading displaced muon with $p_T(\mu_1) >  200$ GeV.

%%%%%%%%%%%%%%%%%%%%%%%%%%%%%%%%%%%%%%%%%%%%%%%%%%%%%%%%%%%%%%%%%%%%%%%%%%%%%%%%%
\subsection{Simulation details}
\label{sec:sim}
%%%%%%%%%%%%%%%%%%%%%%%%%%%%%%%%%%%%%%%%%%%%%%%%%%%%%%%%%%%%%%%%%%%%%%%%%%%%%%%%%

To analyse the kinematics and later on sensitivity, we simulate the signal events using the following steps. We use the Universal FeynRules Output (UFO)~\cite{Degrande:2011ua} of $B-L$ model developed in Ref.~\cite{Deppisch:2018eth} in combination with the Monte Carlo event generator {\tt MadGraph5aMC$@$NLO} -v2.6.7~\cite{Alwall:2014hca} at parton level. The FeynRules~\cite{Alloul:2013bka,Christensen:2008py} model file and UFO is publicly available from the FeynRules Model Database at~\cite{FeynrulesDatabase}.  For every signal sample, we generate $10^4$ signal events. We then pass the generated parton level events on to {\tt PYTHIA v8.235}~\cite{Sjostrand:2014zea} which handles the initial and final state parton shower, hadronization, heavy hadron decays, etc. The clustering of the events is performed by {\tt FastJet v3.2.1}~\cite{Cacciari:2011ma}\footnote{When analysing the hadron level events, we remove the muons from the $N$ decays in the genjets collection.}. Additionally, we use the \verb|NNPDF23_lo_as_0130_qed| PDF set. Using the same setup we also produce $10^5$ background events in each $t\bar{t}$ (leptonic), $ZWW$ (leptonic), and $\mu\nu Z$ (leptonic) final states. In particular, we consider only muon flavoured final states for backgrounds, which will be compatible with our signal final states.

Finally in order to minimise the Monte-Carlo efforts, heavy neutrinos are always decayed promptly. We simulate displacement via inverse sampling of the decay distribution. Specifically, the efficiency of our displaced vertex analysis relies on both the geometrical acceptance of the detector, $\epsilon_\text{geo}$, as well as the reconstruction effects, $\epsilon_\text{recon}$, and the total efficiency is  $\epsilon_\text{DV} = \epsilon_\text{geo} \times \epsilon_\text{recon}$. The geometrical acceptance of the detector includes the probability of the heavy neutrino travelling a distance $L$ from the interaction point before decaying, given by the exponential density distribution
\begin{align}
	p(L) dL = \frac{dL}{L_N} e^{-L/L_N},
\end{align} 
where $L_N$ is the decay length of the boosted heavy neutrino in the laboratory frame. The efficiency $\epsilon_\text{geo}$ is calculated by inverse sampling of the cumulative decay length distribution function, which in turn depends on the heavy neutrino mass and boost, and the active-sterile mixing. The reconstruction efficiency is assumed $\epsilon_{recon} = 100~\%$ unless otherwise stated. 

%%%%%%%%%%%%%%%%%%%%%%%%%%%%%%%%%%%%%%%%%%%%%%%%%%%%%%%%%%%%%%%%%%%%%%%%%%%%%%%%%
\section{FCC-hh reach for heavy $Z'$ production}
\label{sec:zpreach}
%%%%%%%%%%%%%%%%%%%%%%%%%%%%%%%%%%%%%%%%%%%%%%%%%%%%%%%%%%%%%%%%%%%%%%%%%%%%%%%%%
%
\begin{figure}[t!]
\centering
\includegraphics[width=0.49\textwidth]{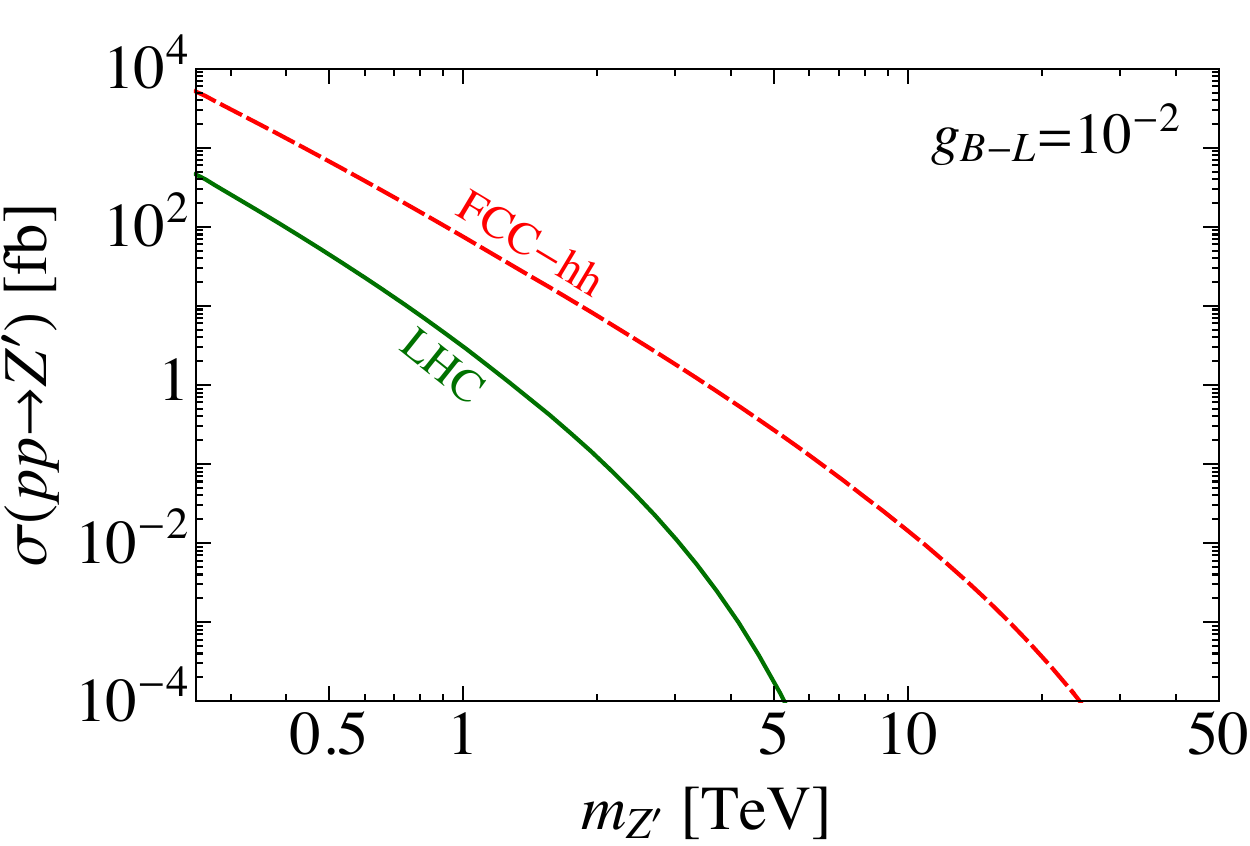}
\includegraphics[width=0.49\textwidth]{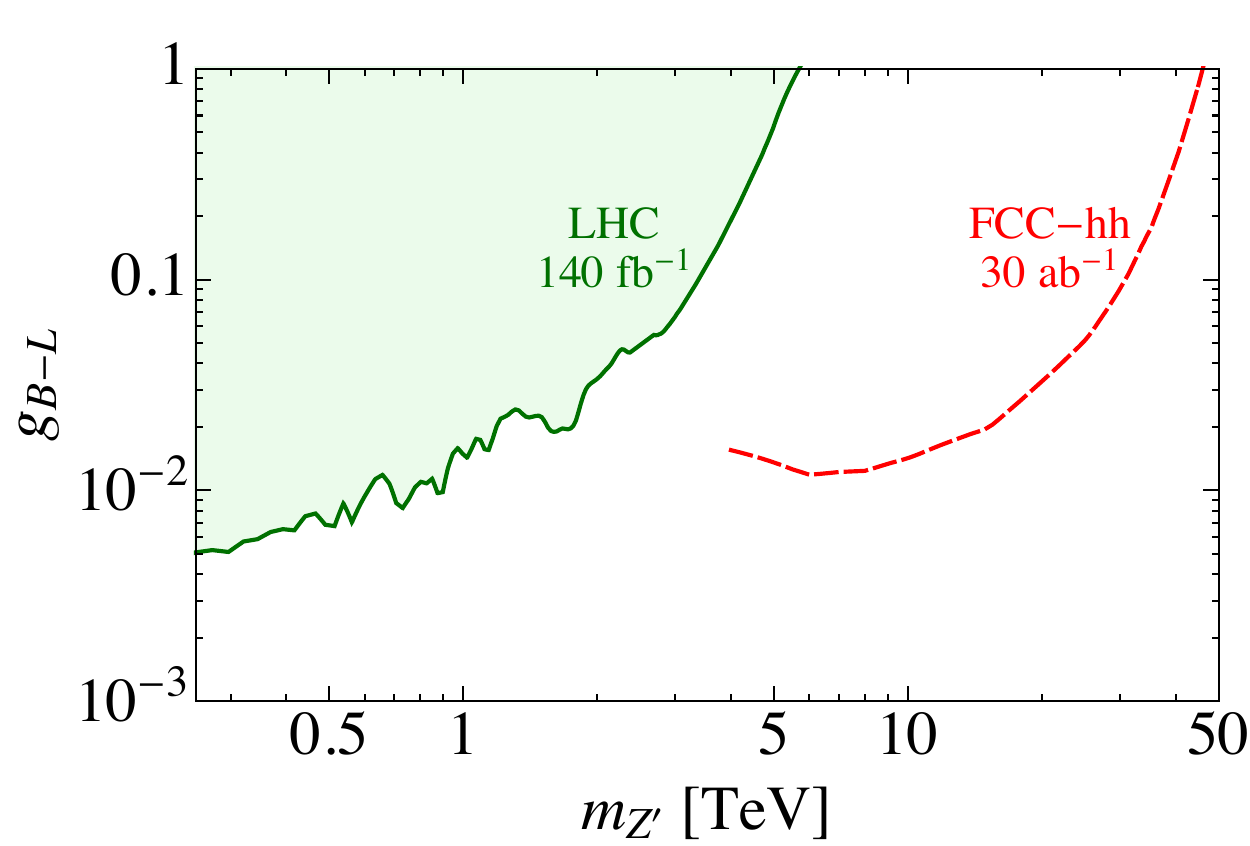}
\caption{Left: Cross section $\sigma(pp\to Z')$ as a function of the $Z'$ gauge boson mass $m_{Z'}$ at the 13 TeV LHC and 100 TeV FCC-hh with $g_{B-L} = 10^{-2}$.
Right: Upper limit on the $U(1)_{B-L}$ gauge coupling $g_{B-L}$ as a function of $m_{Z'}$ at the LHC with 140~fb$^{-1}$ luminosity (green) \cite{CMS:2019tbu,CMS:2021ctt} and projected sensitivity at the FCC-hh with 30~ab$^{-1}$ (red dashed), recast from~\cite{Helsens:2019bfw}. }
\label{fig:zp_limits}
\end{figure}
Before we proceed, we estimate the FCC and LHC potential to probe $Z'$ resonance mass. A detailed investigation of low mass $Z'$ production mechanisms and LHC limits was carried out in~\cite{Deppisch:2019ldi} and \cite{Amrith:2018yfb}. In this study, we are instead interested in high mass $Z'$ and the associated experimental sensitivity. To this end we exploit the existing CMS resonance search in dilepton final state at 140~fb$^{-1}$ luminosity~\cite{CMS:2019tbu}. Comparable results can be obtained by recasting ATLAS search in the same final state~\cite{Aad:2019fac}. We furthermore compliment these limits with the projected FCC-hh limits as detailed in~\cite{Helsens:2019bfw}. We recast both these limits for the $B-L$ coupling. The recast is performed as follows. We calculate the simulated cross section $\sigma^{\rm{ref}}(pp \to Z' \to l\bar{l})$  using the $B-L$ model UFO with fixed $g_{B-L}^{\text{ref}}$ without implementing any generator cuts, then we require the cross section $\sigma(\text{U.L.})$ to be equal to the upper limits from the Ref.~\cite{CMS:2019tbu}, therefore $(g_{B-L}^{\text{U.L.}})^2 / (g_{B-L}^{\text{ref}})^2 \times \sigma^{\rm{ref}}(pp \to Z' \to l\bar{l})\sim  \sigma(\text{U.L.})$.

In Fig.~\ref{fig:zp_limits}, we show the resulting $Z'$ production cross section (left panel) and the limits on the $g_{B-L}$ coupling (right panel) at HL-LHC (green) and FCC-hh (red), respectively. Comparing the $Z'$ production cross section (left panel) between HL-LHC and FCC-hh, it is clear that FCC-hh will gain significant cross section for the production of heavier $Z'$, thus extending the reach for heavier resonances.

This gain in cross section is complemented by a large luminosity of 30 ab$^{-1}$, and they are in turn reflected in the more stringent $g_{B-L}$ limits (right panel). Starting from about $m_{Z'} = 4$~TeV the FCC-hh will improve the $g_{B-L}$ limit by at least an order of magnitude. The lack of significant gain at low $Z'$ masses is due to the large dilepton background at the FCC-hh~\cite{Helsens:2019bfw}. This background becomes much smaller at high mass, and correspondingly high mass resonances can be probed. For $Z'$ lighter than $\sim 4$~TeV, the $Z'$ decays to heavy neutrinos can lead to displaced final states and may be beneficial for exploring lower $Z'$ mass due to smaller backgrounds. We will not explore this region any further however such a strategy might be worth exploring further.

It is thus clear that FCC-hh will in general be able to produce much heavier $Z'$ as compared to HL-LHC and correspondingly improve the limits on $g_{B-L}$. This has important implications for the heavy neutrinos produced via $Z'$. First and foremost, the heavy neutrino will in general be produced with a large boost which will produce more energetic decay products in the final state. Second, such large boosts will result in longer decay lengths of the heavy neutrinos producing more displaced objects in the detector.

We require $\sigma(pp \to Z') > 10^{-2}$~fb in order to achieve $\sim 10$ signal events  $Z'$ $\to N N$ final state at 30 ab$^{-1}$. Considering the FCC-hh's reach as shown in Fig.~\ref{fig:zp_limits}, we assume a benchmark point of $m_{Z'} = 5$~TeV and $g_{B-L} = 10^{-2}$ in the rest of the paper.

%%%%%%%%%%%%%%%%%%%%%%%%%%%%%%%%%%%%%%%%%%%%%%%%%%%%%%%%%%%%%%%%%%%%%%%%%%%%%%%%%
\section{Heavy neutrino production at the FCC-hh}
\label{sec:HPD}
%%%%%%%%%%%%%%%%%%%%%%%%%%%%%%%%%%%%%%%%%%%%%%%%%%%%%%%%%%%%%%%%%%%%%%%%%%%%%%%%%

Having understood the reach of the FCC-hh in terms of the $Z'$ mass and gauge coupling, we are now equipped to compute the heavy neutrino production cross section and decay modes. To this end, we explore the heavy neutrino production via the SM $W,Z$ mediators and via the $B-L$ $Z'$. We furthermore also demonstrate effect of $p_T$ cut on leading lepton, in accordance with our trigger strategy. We simulate these processes as discussed in section~\ref{sec:sim}.

%%%%%%%%%%%%%%%%%%%%%%%%%%%%%%%%%%%%%%%%%%%%%%%%%%%%%%%%%%%%%%%%%%%%%%%%%%%%%%%%%
\subsection{Heavy neutrino production via SM $W$ boson}
%%%%%%%%%%%%%%%%%%%%%%%%%%%%%%%%%%%%%%%%%%%%%%%%%%%%%%%%%%%%%%%%%%%%%%%%%%%%%%%%%

The production of heavy neutrinos from SM $W, Z$ bosons will always remain primary mechanism as it only relies on the active-sterile mixing naturally expected to generate neutrino masses. In this section we will illustrate the heavy neutrino production cross sections from both $W$ and $Z$ mediators, however for our final analysis we will consider only the $W$ mediated process and use the $\mu\mu\nu$ or $\mu j j$ final state from the decay of the heavy neutrinos to assess the sensitivity. Among the lepton final states, only muon flavour is considered as it is cleaner and in general leads to higher detector efficiency compared to electrons or taus.

Benefiting from the 100 TeV collision energy, the cross section of the Drell-Yan process $pp \to W^{\pm} \to \mu^{\pm} \nu$ will increase from $1.56\times 10^{7}$~fb at the 13~TeV LHC to $1.05\times 10^{8}$~fb at the 100~TeV FCC-hh.
\begin{figure}[t!]
\centering
\includegraphics[width=0.49\textwidth]{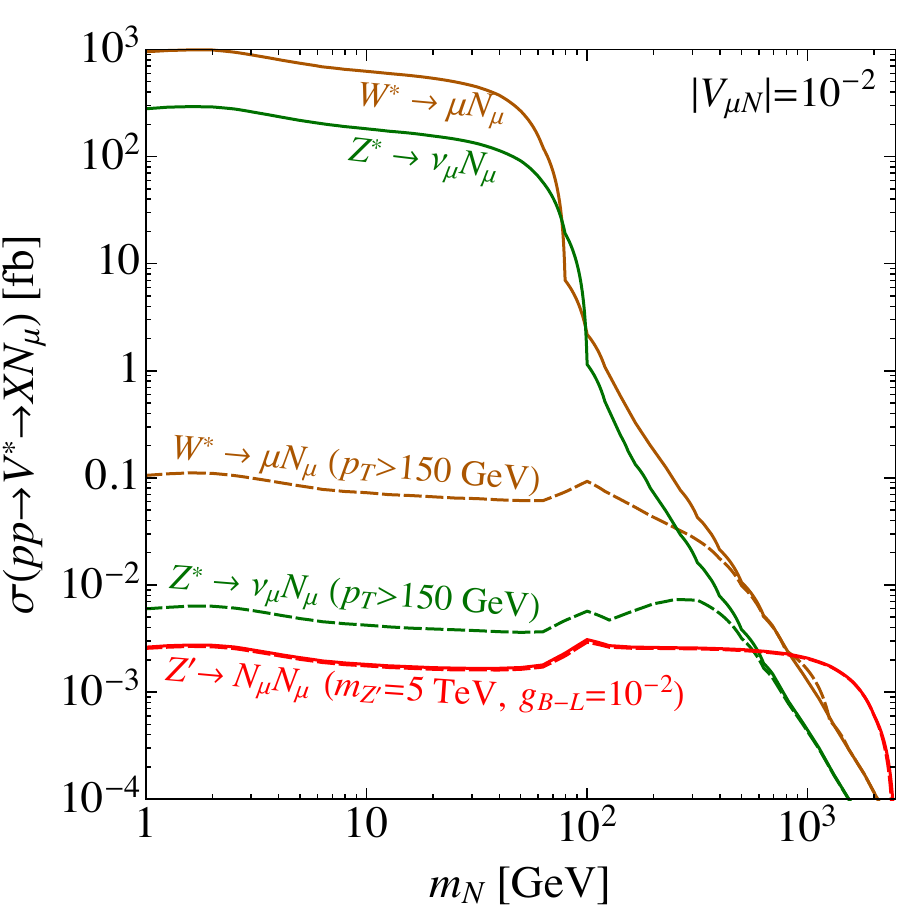}
\includegraphics[width=0.49\textwidth]{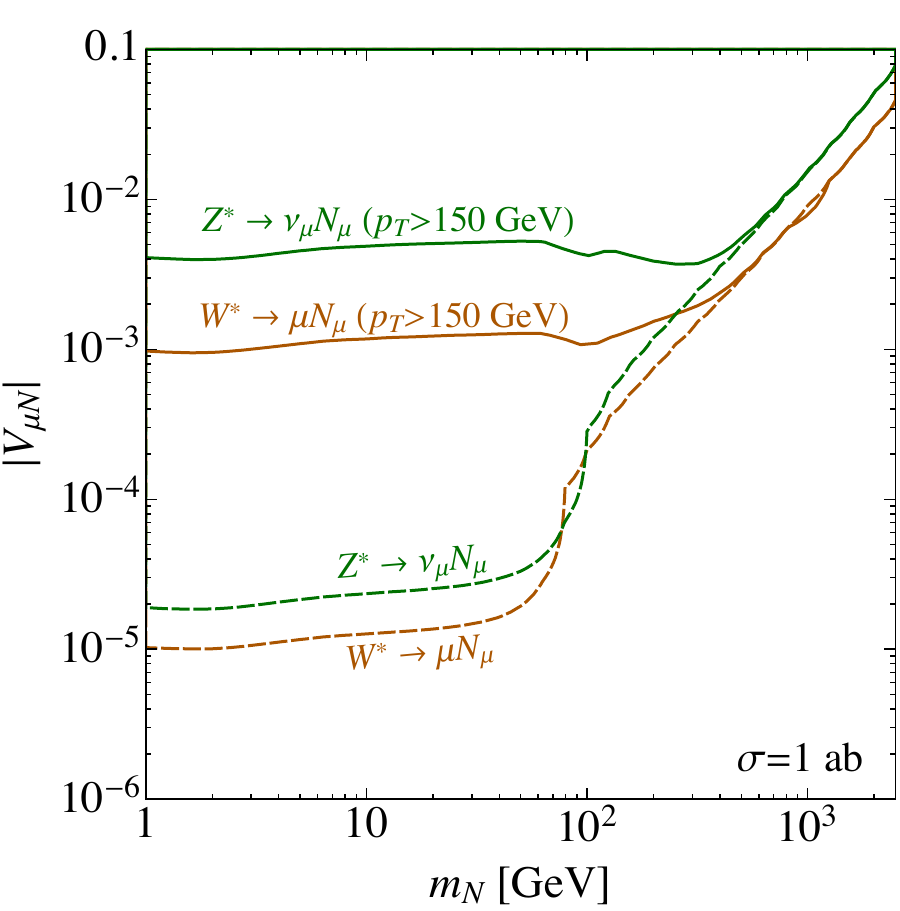}
\caption{Left: Production cross sections of $pp\to W^{\pm} \to N l^{\pm}, N \to \mu^+\mu^- \nu$ (orange), $pp \to Z \to N \nu, N \to \mu^+ \mu^- \nu$ (green) and $pp \to Z' \to NN, N\to\mu^+\mu^-\nu$ (red) as functions of $m_N$ at the 100~TeV FCC-hh. The solid curves are before cuts and the dashed curves are after applying the leading $p_T(\mu) > 150$~GeV cut. The active-sterile mixing is fixed at $|V_{\mu N}| = 10^{-2}$. Right: Contours of heavy neutrino production cross section at 1~ab in the ($m_N$, $|V_{\mu N}|$) plane at the 100~TeV FCC-hh. The curves correspond to SM production channels before (solid) and after (dashed) the $p_T$ cut on final state leptons.}
\label{fig:heavy neutrino_xsecs}
\end{figure}
%

%%%%%%%%%%%%%%%%%%%%%%%%%%%%%%%%%%%%%%%%%%%%%%%%%%%%%%%%%%%%%%%%%%%%%%%%%%%%%%%%%
\subsection{Heavy neutrino production via $Z'$ boson}
%%%%%%%%%%%%%%%%%%%%%%%%%%%%%%%%%%%%%%%%%%%%%%%%%%%%%%%%%%%%%%%%%%%%%%%%%%%%%%%%%

In the case of heavy neutrino production via a $Z'$ mediator, the production cross section is given by $\sigma(pp \to Z^{\prime})\times BR(Z^{\prime}\to N N ) $~\cite{Deppisch:2019kvs}, where the branching ratio is independent of the active-sterile mixing.

Assuming a 5 TeV $Z'$ gauge boson, with $g_{B-L}\sim 10^{-2}$, which saturates the FCC-hh sensitivity and $m_{N}\sim 10 $ GeV, the $pp \to Z' \to N N$ cross section reaches $\sim 10^{-2}$ fb. This cross section will be approximately constant up to $m_N = 2.5$ TeV, beyond which it will vanish due to phase space considerations.

In order to understand the complementarity of $B-L$ heavy neutrino production with the SM mediated production processes, we compare the two cross sections before and after imposing the cut $p_T(\mu_1) > 150$ GeV, consistent with the trigger requirement. In Fig.~\ref{fig:heavy neutrino_xsecs}~(left), we show the heavy neutrino production cross sections via SM $W$, $Z$, as well as $B-L$ $Z'$ with $ N \to \mu^+ \mu^- \nu$ for $|V_{\mu N}| = 10^{-2}$, $g_{B-L}=10^{-2}$ and $m_{Z'} = 5$~TeV (left panel). For the $Z'$ mediated process, only one of the two heavy neutrinos decays. We also show contours of 1 ab production cross section as a function of heavy neutrino mass and mixing angle (right panel).

Several features of these plots are to be noticed: Firstly, the production cross sections via SM mediators should be rescaled by the ratio of the square of the mixing angles for $|V_{\mu\,N}| \neq 10^{-2}$. Such rescaling does not apply for the $Z'$ channel. Therefore, although for $|V_{\mu\,N}| = 10^{-2}$, the SM mediated production cross section is much larger than the $Z'$ counterpart, for smaller mixing angles the situation will be reversed. The cross-over between $Z'$ and SM $W$ mediated channels takes place for $|V_{\mu N}| \sim 10^{-5}$ without the $p_T$ cut, and for $|V_{\mu N}| \sim 10^{-3}$ after the cut. Secondly, the effect of the $p_T$ cut is much stronger on the SM mediated channel, while for the $Z'$ mediated process there is virtually no effect of the $p_T$ cut. This is because the heavy $Z'$ mass already leads to energetic displaced leptons in the final state, the lighter $W$ and $Z$ on the contrary lead to softer prompt leptons. As can be seen from the plot, the cross section via SM mediators can drop by three order magnitude after such a lepton $p_T$ cut. Finally, the SM $Z$ mediated cross section is approximately one order magnitude smaller than the corresponding $W$ mediated process, which justifies our decision to neglect the $Z$ mediated processes. 

%%%%%%%%%%%%%%%%%%%%%%%%%%%%%%%%%%%%%%%%%%%%%%%%%%%%%%%%%%%%%%%%%%%%%%%%%%%%%%%%%
\subsection{Kinematics of the heavy neutrino and its final states}
\label{sec:kinematics}
%%%%%%%%%%%%%%%%%%%%%%%%%%%%%%%%%%%%%%%%%%%%%%%%%%%%%%%%%%%%%%%%%%%%%%%%%%%%%%%%%

Having explained the heavy neutrino production cross sections in the previous section, we also discuss the kinematics of the final states by considering the two production channels of the heavy neutrinos either via a SM $W$ or $B-L$ $Z'$. For the final states, we only take $N \to \mu^+ \mu^- \nu$ as an example. We compare the $p_T$ distributions of the resulting heavy neutrino, muons, and $W^{(*)}$ to illustrate advantages of each of the production modes. 

\begin{figure}[t!]
\centering
\includegraphics[width=0.49\textwidth]{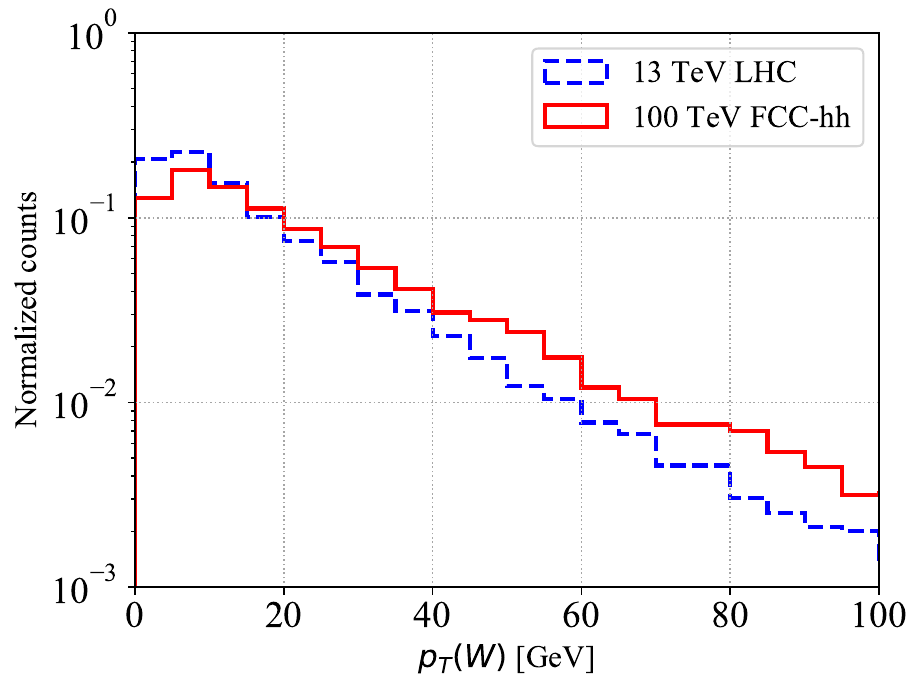}
\includegraphics[width=0.49\textwidth]{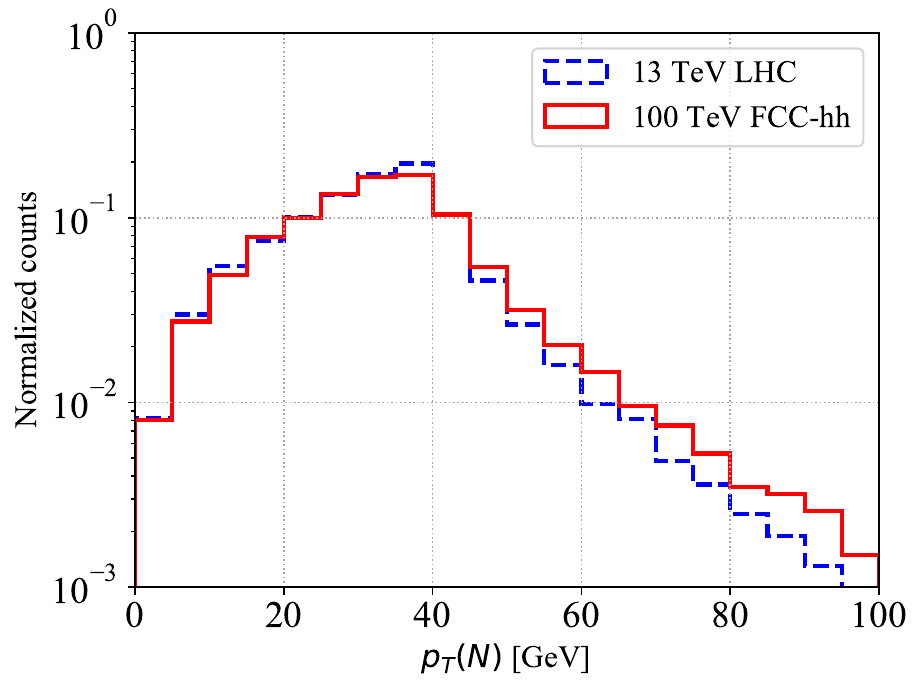}
\includegraphics[width=0.49\textwidth]{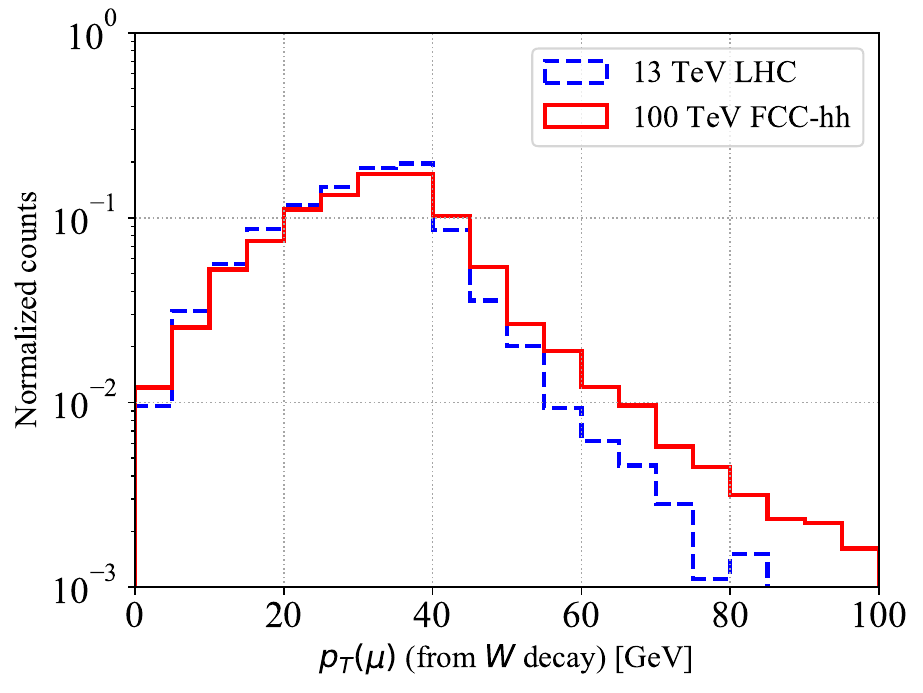}
\includegraphics[width=0.49\textwidth]{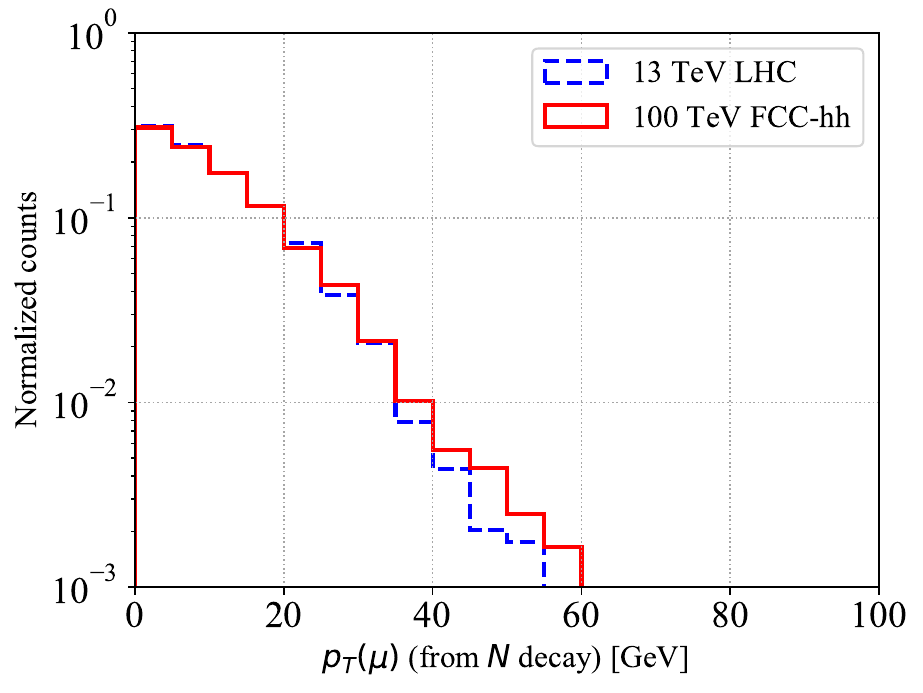}
\caption{Transverse momentum distribution of particles in the process $pp \to W^{\pm} \to N \mu^{\pm}, N \to \mu \mu \nu$:
$p_T(W)$~(upper left panel), $p_T(N)$~(upper right panel), $p_T(\mu)$ from the $W$ boson decay~(bottom left panel) and $p_T(\mu)$ from the $N$ decay~(bottom right panel) at the 13 TeV LHC (blue dashed) and 100 TeV FCC-hh (red solid). }
\label{fig:ptW}
\end{figure}

%%%%%%%%%%%%%%%%%%%%%%%%%%%%%%%%%%%%%%%%%%%%%%%%%%%%%%%%%%%%%%%%%%%%%%%%%%%%%%%%%
For the SM mediated production, we compare the kinematics between LHC and FCC-hh center of mass energies. In Fig.~\ref{fig:ptW} (top row), we plot the $p_T$ of the mediator $W^{(*)}$ (left panel) and $N$ (right panel) in $p p \to W^{(*)} \to N \mu$ at the 13 TeV LHC (blue dashed) and FCC-hh (red solid). As the $W^{(*)}$ boson is produced in s-channel, it has a small transverse momentum. In the left panel, with larger energy, the average of the $p_{T}(W^{(*)})$ increases by a few GeV at the FCC-hh, while most of the $W^{(*)}$ bosons still possess vanishing $p_T$. Correspondingly, as shown in Fig.~\ref{fig:ptW}~(right) the heavy neutrino $p_T$ does not increase significantly at the FCC-hh as compared to the LHC. The gain in sensitivity at the FCC-hh is therefore primarily due to the larger production cross section.

In Fig.~\ref{fig:ptW} (bottom row), we plot the $p_T$ of the (prompt) muon from $W^{(*)}$ and from $N$ decays. We fix the mass of the heavy neutrino to 10 GeV for concreteness. As the $W^{(*)}$ or the $N$ do not gain significant energy at the FCC-hh as compared to the LHC, the resulting final states also do not gain any more $p_T$, as reflected in the muon $p_T$ distributions shown. As we will see later, for this reason, FCC-hh does not lead to a large gain in sensitivity for SM mediated heavy neutrino production.
%

\begin{figure}[t!]
\centering
\includegraphics[width=0.49\textwidth]{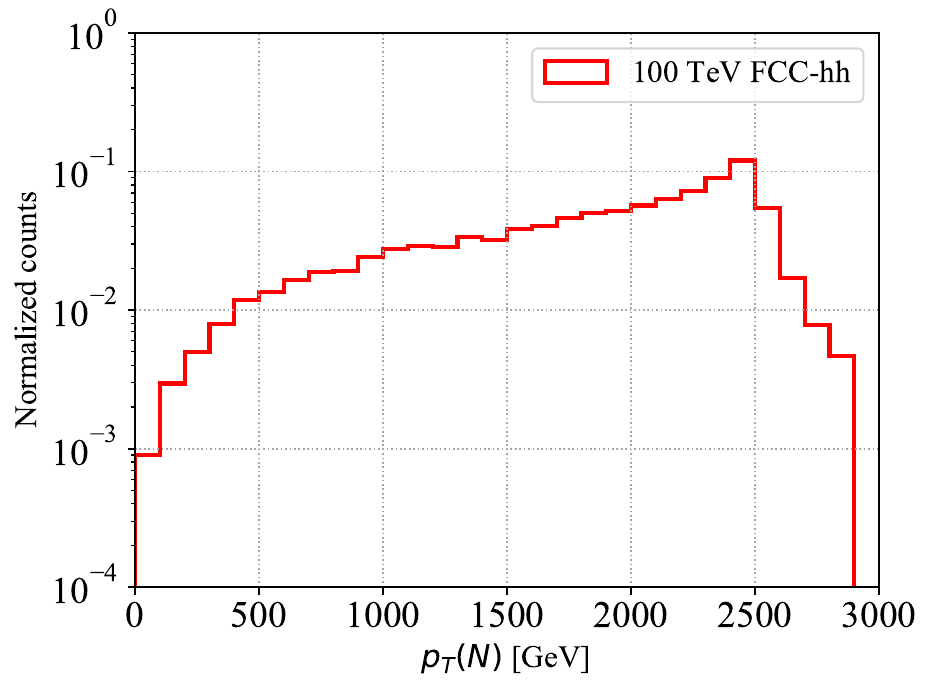}
\includegraphics[width=0.49\textwidth]{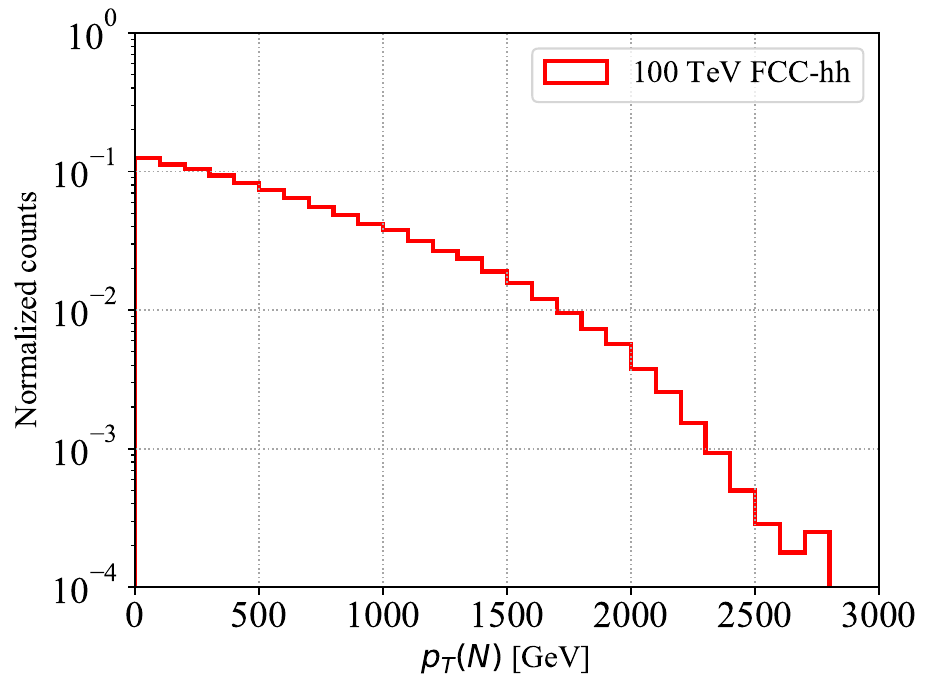}
\caption{Left Panel: $p_T(N)$ in the $pp \to Z' \to NN$ process at the FCC-hh. Right Panel: $p_T(\mu)$ from the $N$ decay.}
\label{fig:BLpt}
\end{figure}

%%%%%%%%%%%%%%%%%%%%%%%%%%%%%%%%%%%%%%%%%%%%%%%%%%%%%%%%%%%%%%%%%%%%%%%%%%%%%%%%%
\label{sec:BL}
%%%%%%%%%%%%%%%%%%%%%%%%%%%%%%%%%%%%%%%%%%%%%%%%%%%%%%%%%%%%%%%%%%%%%%%%%%%%%%%%%
In contrast to the SM production, the heavy neutrinos in $B-L$ production can have significant $p_T$, as they are produced from a heavy resonance, i.e., a 5 TeV $Z'$ (as shown in Fig.~\ref{fig:BLpt} left). Therefore, in $N\to \mu \mu \nu$ decay, the final $\mu$ can have $p_T \sim \mathcal{O}$(100) GeV (see Fig.~\ref{fig:BLpt} right). It thus becomes easier to pass the stringent $p_T$ requirements e.g. for trigger purposes. However, due to the presence of the heavy $Z'$, it is also possible that the decay products of the heavy neutrino are collimated. 

\begin{figure}[t!]
\centering
\includegraphics[width=0.49\textwidth]{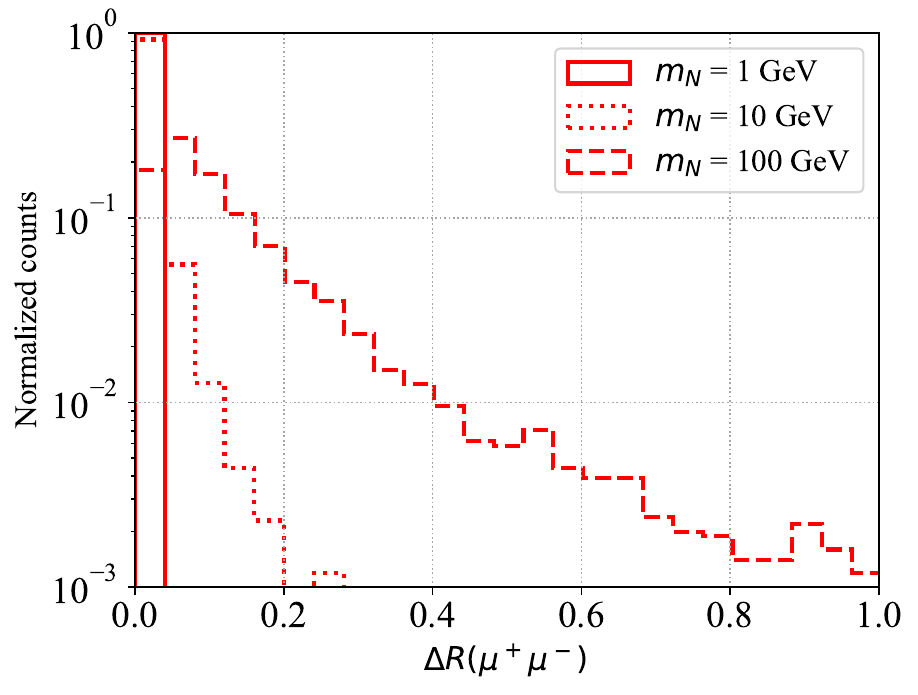}
\includegraphics[width=0.49\textwidth]{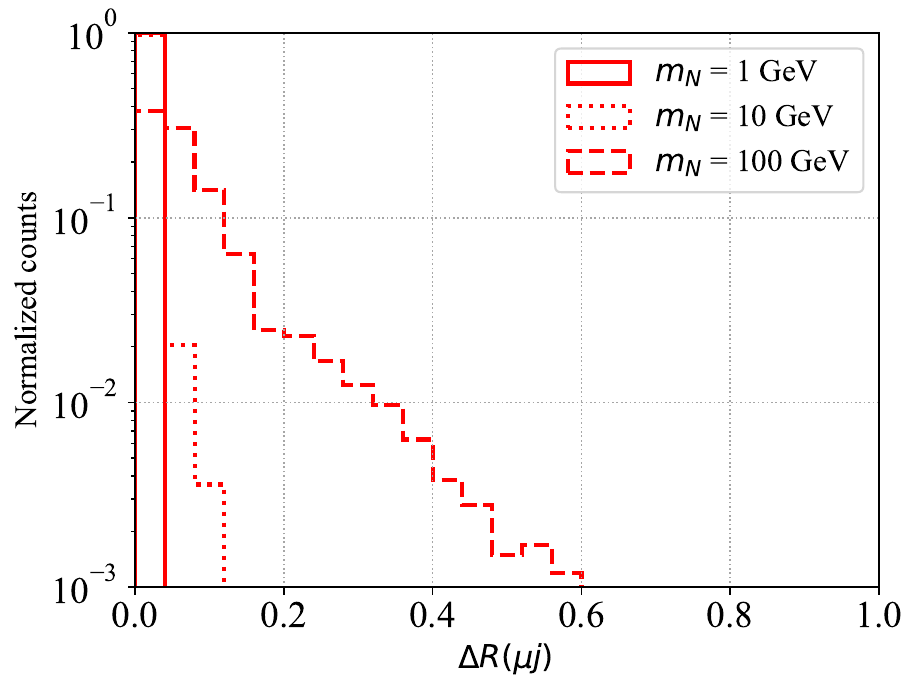}
\caption{Left: $\Delta R (\mu^+ \mu^-)$ from $pp\to Z' \to N N, N \to \mu \mu \nu$ for $m_N =$ 1, 10, 100 GeV. Right: $\Delta R (\mu j)$ from the $pp\to Z' \to N N, N \to \mu j j$ for $m_N =$ 1, 10, 100 GeV. $m_{Z'}$ is fixed at 5 TeV.  All distributions are shown for FCC-hh.}
\label{fig:deltaR}
\end{figure}

To illustrate this, in Fig.~\ref{fig:deltaR} (left), we plot the $\Delta R$ between the final state muons~(and jet) for the leptonic $N$ decay ($N\to \mu \mu \nu$) for three heavy neutrino masses of 1~GeV (solid line), 10~GeV (dotted line) and 100~GeV (dashed line). We analyse here decays of only one of the heavy neutrinos. In principle, larger muon multiplicities can be obtained if decays of the second heavy neutrino are also considered. Such muons coming from two different heavy neutrinos will however not lead to a vertex and hence can be discarded. We therefore assume that only one heavy neutrino decays to illustrate potentially non-isolated muons. From these distributions it is clear that muons can be very collimated when the mass difference between $Z'$ and heavy neutrino is large. To demonstrate a similar situation in the semi-leptonic decay ($N\to \mu j j$), in Fig.~\ref{fig:deltaR} (right) we plot the minimum $\Delta R$ between the muon and two of the jets emerging from $N\to \mu j j$. In order to remove any soft hadronic activity in the event we consider jets with $p_T > 100 $ GeV. It can be seen that for heavy neutrinos lighter than 10 GeV, the $\Delta R$ is smaller than 0.2. This poses potential isolation problems in leptonic final states and can also lead to fat jets in hadronic final states. 
%

%%%%%%%%%%%%%%%%%%%%%%%%%%%%%%%%%%%%%%%%%%%%%%%%%%%%%%%%%%%%%%%%%%%%%%%%%%%%%%%%%
\section{Sensitivity}
\label{sec:sensitivity}
%%%%%%%%%%%%%%%%%%%%%%%%%%%%%%%%%%%%%%%%%%%%%%%%%%%%%%%%%%%%%%%%%%%%%%%%%%%%%%%%%
We now estimate the sensitivity of the FCC-hh with 30 ab$^{-1}$ integrated luminosity to heavy neutrino production both via SM $W$ and $B-L$ $Z'$ mediators. For each of these production modes we consider two different heavy neutrino decay channels, namely $\mu\mu\nu$ and $\mu jj$, which correspond to either the leptonic or hadronic $W$ decay. For the SM $W$ mediated mode, we always use the prompt lepton as a trigger. For the $B-L$ $Z'$ mode, we use the prompt and the displaced lepton triggers. We furthermore consider any RH neutrino decaying within the detector volume as outlined in Section~\ref{sec:FCC} and $L_N > 1$~mm as displaced.

\begin{table}[t!]
\centering
\begin{tabular}{|c|c|c|c|c|} 
\hline
\multirow{2}{*} & 3$\mu + \slashed{E_T}$ & $2\mu + 2j$ & $2\mu + 4j$ & $4\mu + \slashed{E_T}$\\ 
& & (OS/SS) & (OS/SS)  &  \\ 
\hline
\multirow{4}{*} {Common cuts} & \multicolumn{4}{|c|}{$p_T(\mu_1) > 150\,\rm{GeV}$,} \\
& \multicolumn{4}{|c|}{$p_T(\mu) > 20\,\rm{GeV}$} \\
& \multicolumn{4}{|c|}{$p_T(j) > 20\,\rm{GeV}, $} \\
& \multicolumn{4}{|c|}{$ |\eta(\mu,j)| \leq 4$} \\
\hline
Lepton charge & N.A. & $\mu^+\mu^-$/$\mu^{\pm}\mu^{\pm}$ & $\mu^+\mu^-$/$\mu^{\pm}\mu^{\pm}$ & N.A.\\
\hline 
Number of light jets &  0 & $ = 2$ & $= 4$ & 0 \\
\hline 
b-jet veto &  No & Yes & Yes &  No \\
\hline
$\cancel{\it{E}}_{T}$  & N.A. & $< 20\,\rm{GeV}$ & $< 20\,\rm{GeV}$ & $ > 100\,\rm{GeV}$ \\
\hline
$M_{inv}$ & N.A. & $> 4\, \rm{TeV}$& $> 4\, \rm{TeV} $ & N.A. \\
\hline
\end{tabular}
\caption{Different cuts on final states for various prompt signal categories considered.} \label{tab:sig_cuts}
\end{table}
Before we calculate the sensitivity, an estimation of the background is necessary. In general, this analysis can be carried out in several categories depending on charge and flavour composition of the signal processes (see~\cite{Antusch:2016ejd} for such an exercise for SM mediated heavy neutrino production). We however restrict ourselves to muons in the final state and either consider same sign or opposite sign muons when a sufficient number of signal events are available. Unsurprisingly, the charge composition is relevant only when the signal originates from semi-leptonic decays of a heavy neutrino ($\mu jj$ final state). As indicated before, the main signal and background discriminating variables we use are missing energy, number of jets, lepton charge composition and invariant mass of the system. These are summarised in Table~\ref{tab:sig_cuts}. In Table~\ref{tab:bck}, we show the main background processes and approximate number of expected events for a luminosity of 30 ab$^{-1}$ after respective cuts.

We begin by discussing background composition and reduction methodologies for SM mediated heavy neutrino production and prompt decays. For heavy neutrino decays to muonic final states ($\mu\mu\mu\nu$), the background consists of $ \mu^\pm \nu Z$ with leptonic $Z$ decays. This background has approximately $10^5$ events, therefore we require $\sim 10^3$ signal events. It should be noted that we attempt no optimization in terms of final state flavour composition. For the hadronic final state, ($\mu\mu jj$), $t\bar{t}$ backgrounds are dominant. As the signal contains a fully visible final state, the resulting missing energy is small. Therefore, missing energy is a relevant discriminator here. For the opposite sign muons category requiring $p_T(\mu_1) > 150$~GeV, $\cancel{\it{E}}_T < 20$~GeV and putting a b-jet veto on the leptonic $t\bar{t}$ decays leads to $\sim 10^4$ events. Requiring same sign leptons (muons) in the final state leads to a more promising situation where backgrounds emerge from charge misidentification. Assuming an optimistic rate of 0.1\% based on LHC detector performance~\cite{ATLAS:2019jvq} leads to $\sim 10$~events~\cite{Antusch:2016ejd}.

For the prompt final states of the $B-L$ mediated processes, two heavy neutrinos are produced. The signal final states we concentrate on are leptonic decays of both heavy neutrinos ($4\mu$ + MET) and semi-leptonic decays of both heavy neutrinos ($2\mu+4j$)~\footnote{The two RH neutrinos can also decay leptonically and semi-leptonically respectively, yielding a $3\mu + 2j + \cancel{E}_T$ final state. The main background comes from $ZWW$ decays similarly, and its cross section is between the $4\mu$ + MET and $2\mu+4j$ final states. We however do not consider this channel in this work as it should not give better sensitivity.}. For $4\mu$ + MET final state, triple boson background processes ($ZWW$) have $\sim10^3$ events. For the $2\mu+4j$ final state, the $t\bar{t}$ mode in opposite sign muon and same sign muon final states are controlled by requiring low missing energy  $\cancel{\it{E}}_T < 20$~GeV and high invariant mass $M(t\bar{t}) > 4$~TeV. For the opposite sign final states, there are still $\sim 10^3$ background events after the cuts, while they become negligible after additionally requiring charge misidentification for the same sign final states.

Finally, for the displaced final states, in the case of SM $W$ mediated production, we always use the available prompt lepton with a $p_T > 150$~GeV to emulate the trigger and require either two displaced muons, in case of $N\to \mu\mu\nu$ decays or one displaced muon and one jet, for $N\to \mu j j $ decays. In addition, we require that the heavy neutrino decays within the detector volume, with geometry as described in Section~\ref{sec:FCC}. We consider the background to be negligible as the signal is sufficiently displaced~\cite{ATLAS:2019kpx, LHCb:2020akw, CMS:2022fut}, and only $N_{\text{signal}} > 3.09$ is required from the Poisson distribution at 95\% confidence level (C.L.). 
\begin{table}[t!]
	\centering
	\begin{tabular}{|l|c|c|c|c|c|}
		\hline
		SM Prompt & Background & $\sigma$(fb) & M($t \bar{t}$) &$N_{\text{B}}$  \\
		\hline
		Leptonic ($\mu\mu\mu \cancel{\it{E}}_{T} $)  &  $ \mu^\pm \nu Z $& $ 11.9$  & - & $3.55 \times 10^5$ \\
		\hline
		Hadronic OS ($\mu^\pm \mu^\mp jj$) & $t \bar{t}$~(leptonic decay)  & 1.84 & - & $5.52 \times$ $10^4$ \\
		Hadronic SS ($\mu^\pm \mu^\pm jj$) & $t \bar{t}$~(leptonic decay)  & 1.84 $\times 10^{-3}$ & - & 55.2 \\
		\hline
		\hline
		$B-L$ Prompt & Background & $\sigma$(fb) & M($t \bar{t}$) &$N_{\text{B}}$  \\
		\hline
		Leptonic ($\mu\mu\mu\mu \cancel{\it{E}}_{T}$)  &  $ Z W W$ & $ 5.92 \times 10^{-2}$  & - & $1.78 \times 10^3$ \\
		\hline
		Hadronic OS ($\mu^\pm \mu^\mp jjjj$) & $t \bar{t}$~(leptonic decay)  & 1.85  &  $8.73 \times 10^{-2}$ & $2.62 \times 10^3$ \\
		Hadronic SS ($\mu^\pm \mu^\pm jj j j$) & $t \bar{t}$~(leptonic decay) & 1.85 $\times 10^{-3}$ & Negligible & Negligible \\
		\hline \hline
	    Displaced Vertex & Background &  $\sigma$(fb) &  M($t \bar{t}$) &$N_{\text{B}}$  \\
		\hline
		Leptonic ($\mu\mu \cancel{\it{E}}_{T}$)  & - & - & - & Negligible  \\
		\hline
		Hadronic ($\mu jj$) & - & - & - & Negligible  \\
		\hline
	\end{tabular}
	\caption{Main background processes and their approximate number of events for the corresponding signal at the FCC-hh with 30 ab$^{-1}$ integrated luminosity. A cut $p_T > 150~(20)$ GeV for the leading~(subleading) muon and $p_T > 20$ GeV for jets are put. The number of leptons and jets are required to match the signal. In addition, $\cancel{\it{E}}_{T}$ > 100 GeV is put for $ZZZ$ background to cut out the ones from $ZZ$.
	And $\cancel{\it{E}}_{T}$ < 20 GeV with b-tag veto with an efficiency of 0.3 for each jet, and M($t \bar{t}$) > 4 TeV are applied for the $t \bar{t}$~\cite{Antusch:2016ejd}.  }
	\label{tab:bck}
\end{table}
\begin{figure}[t!]
\centering
\includegraphics[width=0.99\textwidth]{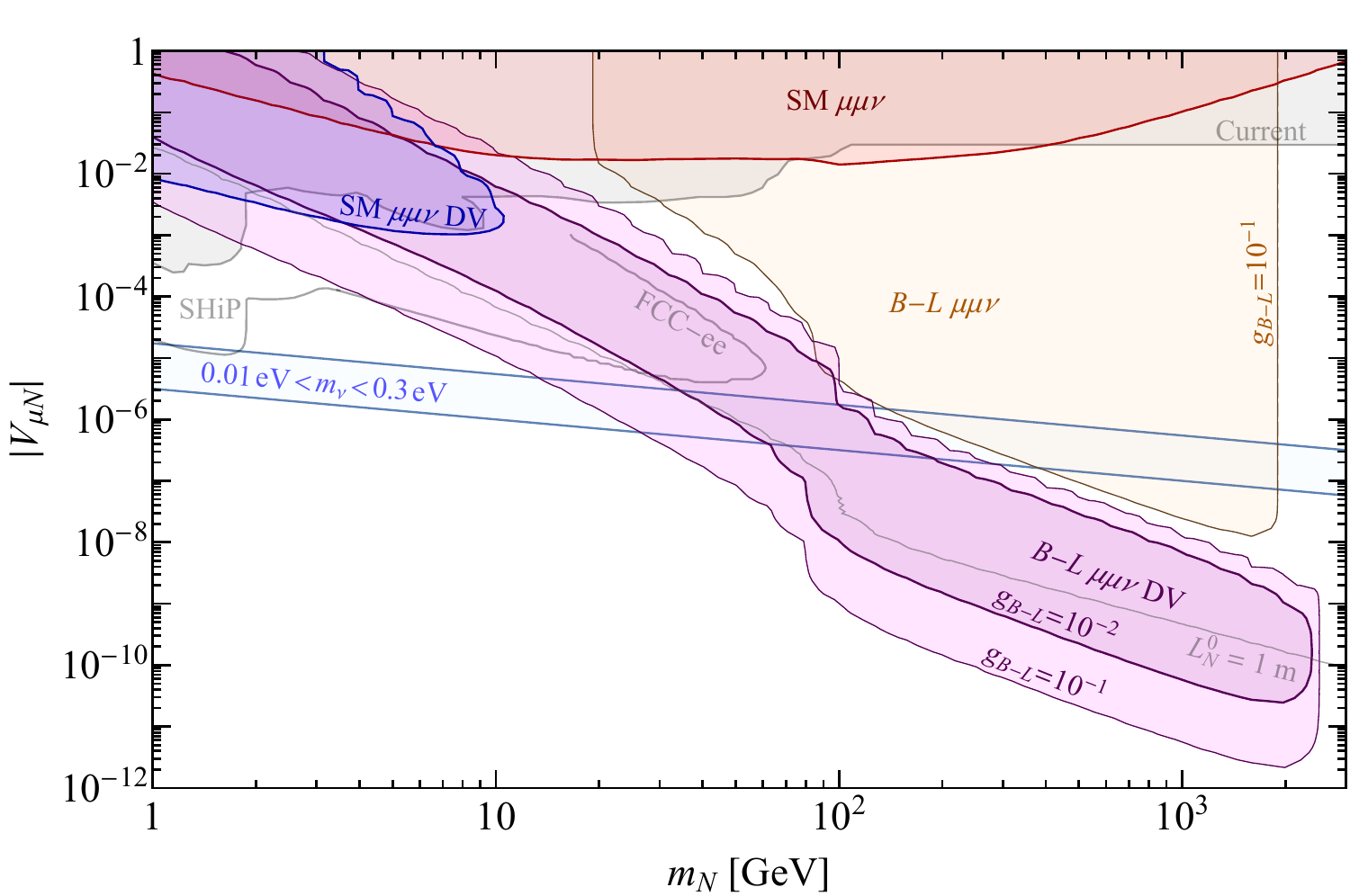}
\caption{Sensitivity of the FCC-hh with 30~ab$^{-1}$ integrated luminosity at 95\% C.L. towards the heavy neutrino production and leptonic $N \to \mu^+ \mu^- \nu$ channel considered: SM $W$ production with prompt (SM $\mu\mu\nu$, red) and displaced (SM $\mu\mu\nu$ DV, blue) vertices; $B-L$ $Z'$ production with prompt ($B-L$  $\mu\mu\nu$, orange) and displaced ($B-L$ $\mu\mu\nu$ DV, purple) vertices. The $U(1)_{B-L}$ parameters are fixed at $m_{Z'} = 5$~TeV and $g_{B-L} = 10^{-2}, 0.1$ as indicated, with prompt $B-L$ $\mu\mu\nu$ not being sensitive at $g_{B-L} = 10^{-2}$. The light blue band corresponds to the regime with light neutrino masses $0.01\text{ eV} < m_\nu < 0.3$~eV via the seesaw mechanism.  The shaded region labelled `Current' is excluded by existing searches for heavy neutrinos~\cite{Bolton:2019pcu}, whereas `SHiP' and `FCC-ee' indicate the projected future sensitivity at the proposed SHiP~\cite{SHiP:2018xqw} and FCC-ee~\cite{Blondel:2014bra}, respectively. The proper heavy neutrino decay length $L_N^0$ of 1~m is indicated by the corresponding curve.}
\label{fig:sen_semilep}
\end{figure}
The resulting sensitivity at 95\% C.L. is shown in Fig.~\ref{fig:sen_semilep}, for the $\mu\mu\nu$ channel. For $Z'$ mediated production, we fix $m_{Z'} = 5$~TeV, while taking the $B-L$ coupling at two representative values, $g_{B-L} = 10^{-2}$ and 0.1, reflecting either a pessimistic view that the $g_{B-L}$ is taken near the projected sensitivity of the FCC-hh or an optimistic view with $g_{B-L}$ large but satisfying the current LHC limits.

Both neutrinos are assumed to decay via $\mu\mu\nu$ in case of prompt $B-L$ production and thus leading to a $4\mu + \cancel{E}_T$ final state. The production channels and decay modes are accordingly labelled, e.g., `SM $\mu\mu\nu$ DV' denotes the SM production mode with $\mu\mu\nu$ final state, in a displaced vertex. Current best limits as collated in \cite{Bolton:2019pcu}, which includes a displaced search at ATLAS \cite{ATLAS:2019kpx} (small bump around $m_N \lesssim 10$~GeV), and projected sensitivities from the proposed SHiP~\cite{SHiP:2018xqw} and FCC-ee~\cite{Blondel:2014bra} are also shown for comparison.

We start with analysing the SM production channels in displaced (blue shaded region) and prompt (red shaded region) signatures. The striking feature for this production channel is the limited FCC-hh sensitivity. As shown in Section~\ref{sec:kinematics}, this is because the $p_T$ of the muons in the final state does not benefit from the larger collision energy. Instead, the stringent $p_T$ requirements, ($p_T(\mu) > 28$~GeV at the LHC \cite{Aaboud:2017buh} compared to $p_T(\mu) > 150$~GeV at the FCC-hh) limit the reach despite the increased cross section. Therefore, although the production cross section of $N$ from this process can reach $\mathcal{O}(100)$~fb for active-sterile mixing $|V_{\mu N}| \sim 10^{-2}$, the FCC-hh fails to probe a significant parameter space as muons in the final states do not have sufficient $p_T$ to pass the cuts. 

Therefore, we obtain a sensitivity $|V_{\mu N}| \gtrsim 10^{-2}$ for the prompt final states of heavy neutrinos from the SM production at the FCC-hh, comparable to the reach of the LHC~\cite{Aad:2019kiz, Sirunyan:2018mtv, Aaij:2014aba}, as indicated by the `Current' region. Although the cross section drops sharply as the $W$ becomes off-shell when $m_N > m_W$, $p_T$ cuts become more efficient and the resulting sensitivity changes smoothly. The requirement of two same sign leptons helps with background control and in general leads to a sensitivity up to one order of magnitude stronger as compared to the opposite-sign signature. We therefore only show the sensitivity for the same-sign signature. For the displaced final states, as we look for heavy neutrinos with longer lifetime situated at lower $|V_{\mu N}|$, the gain due to negligible background is however cancelled out by the reduced cross section, therefore the sensitivity to $|V_{\mu N}|$ becomes stronger by an order of magnitude only leading to $|V_{\mu N}| \gtrsim 10^{-3}$. The sensitivity vanishes for $m_N \gtrsim 10$ GeV, as the displaced final states requires $|V_{\mu N}| \sim 10^{-4}$, making the cross section insufficient to get any sensitivity in such a parameter space. Comparing to the projected limits from SHiP and FCC-hh, the FCC-hh is not competitive at lower $m_N$, but can have and advantage at greater $m_N$ due to its larger collision energy.

\begin{figure}[t!]
	\centering
	\includegraphics[width=0.99\textwidth]{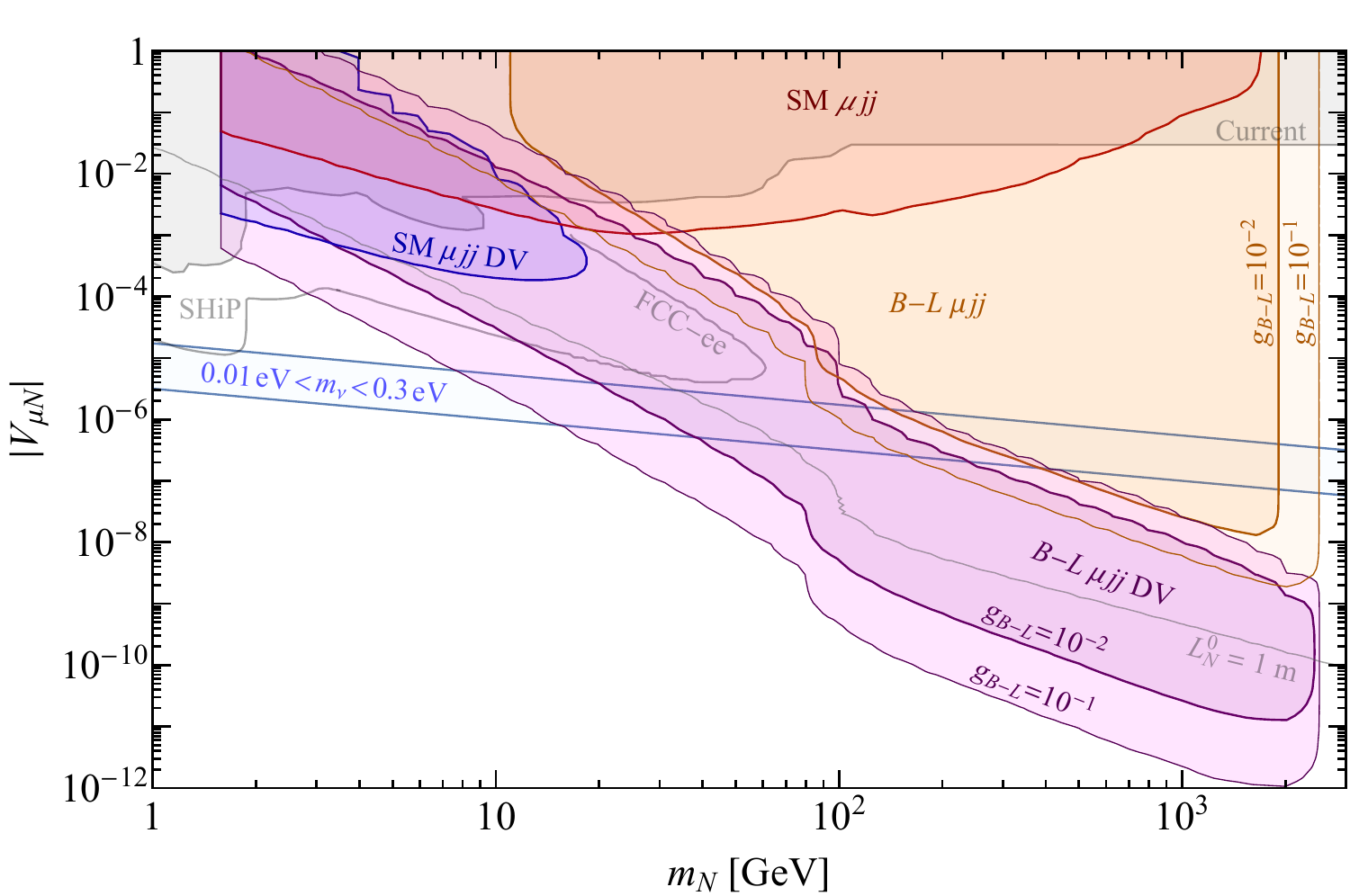}
	\caption{As Fig.~\ref{fig:sen_semilep}, but for the semi-leptonic final state $\mu jj$. }
	\label{fig:sen_hadronic}
\end{figure}
The $B-L$ heavy neutrino production complements the SM channel. In the prompt $\mu\mu\nu$ final state, for $g_{B-L} = 10^{-1}$, the sensitivity extends up to $m_N = 2$~TeV and active-sterile mixing strengths $|V_{\mu N}|$ down to regimes where the heavy neutrino becomes long lived. Because both the neutrino production and decay branching ratios are independent of $|V_{\mu N}|$ ($N$ does not decay to other flavours), only this promptness requirement limits the $|V_{\mu N}|$ reach. For the smaller value $g_{B-L} = 10^{-2}$, no sensitivity is obtained. For the displaced final state, the sensitivity in $V_{\mu N}$ follows heavy neutrino decay lengths between $1~\text{mm}\lesssim L \lesssim 10$~m, extending the $|V_{\mu N}|$ reach by about three orders of magnitude. The break at $m_N \approx 100$~GeV arises due to the change from off-shell to on-shell decays. It is worth noting that combining prompt and displaced final states, the seesaw mechanism responsible for generating light neutrino masses can be tested for $20~\text{GeV} \lesssim m_N \lesssim m_{Z'}/2$, but displaced searches are also sensitive for $g_{B-L} = 10^{-2}$ assuming them being free of background.

The corresponding sensitivity arising from prompt and displaced $\mu jj$ final states is shown in Fig.~\ref{fig:sen_hadronic}. In the prompt $B-L$ channel, both heavy neutrinos are assumed to decay as $N \to \mu jj$ thus leading to a $2\mu + 4j$ final state. We show the prompt final state sensitivity again only for the opposite-sign channel. The most important feature is the higher sensitivity as compared to $\mu\mu\nu$ final states in both displaced and prompt signatures because of the larger branching ratio of the semi-leptonic decay. For the SM $\mu jj$ prompt final state, it is now possible to reach $|V_{\mu N}| \approx 10^{-3}$ for $m_N\approx 30$~GeV. Similar to the SM mediated case, an increased sensitivity in both prompt and displaced final states can be found for $B-L$ heavy neutrino production as well. Otherwise, the overall sensitivity is similar to that of the fully leptonic mode but the prompt $B-L$ $\mu jj$ case is also sensitive for $g_{B-L} = 10^{-2}$ up to masses of $m_N \lesssim 2$~TeV.

\begin{figure}[t!]
\centering
\includegraphics[width=0.99\textwidth]{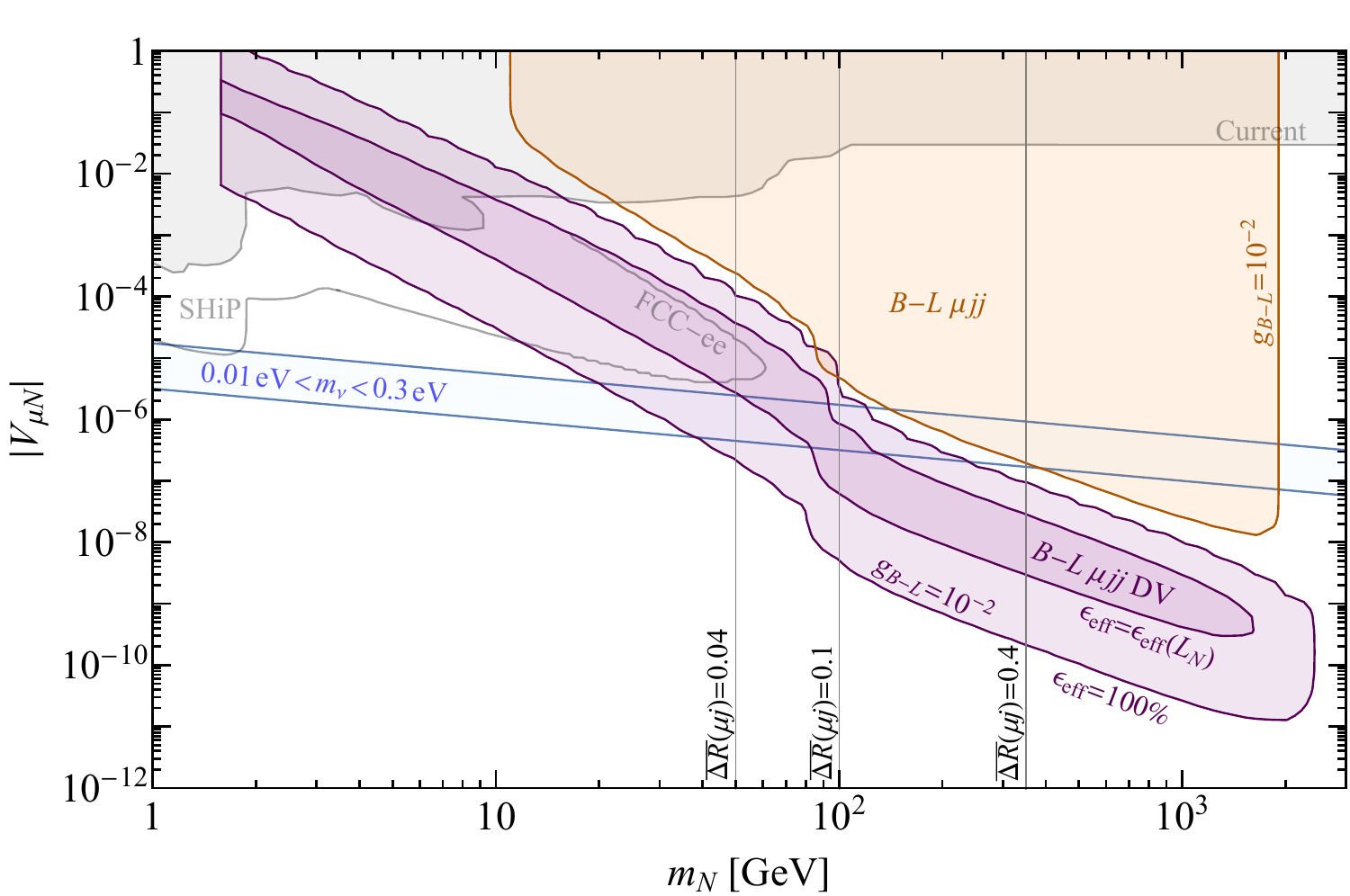}
\caption{As Fig.~\ref{fig:sen_hadronic} but showing the displaced $B-L$ $\mu jj$ sensitivity for both perfect ($\epsilon_\text{eff} = 100\%$) and reduced $\epsilon_\text{eff}(L_N)$ detection efficiency as described in the text. In addition, the vertical lines denote contours of the average $\overline{\Delta R}(\mu j) = 0.04, 0.1, 0.4$ between the muon and closest jet.}
\label{fig:finite_eff}
\end{figure}
We have so far assumed $\epsilon_{recon} = 100$ \% to detect and reconstruct the displaced objects in the detector.
 This is clearly an idealized assumption. In a realistic scenario, the efficiencies will be highest near the primary vertex and will degrade towards outer parts of the detector, see e.g. discussion in~\cite{ATLAS:2019kpx, ATL-PHYS-PUB-2019-013}. To illustrate the impact of less than perfect efficiencies, we show in Fig.~\ref{fig:finite_eff} the sensitivity of the $B-L$ $\mu jj$ displaced final state as in Fig.~\ref{fig:sen_hadronic} for an efficiency dependent on the lab frame decay length $L_N$ of the heavy neutrino,
\begin{align}
\label{lengthapproxi}
	\epsilon_\text{eff}(L_N) = 0.002 \ln^2\left(\frac{L_N}{L_0}\right)\ln^2\left(\frac{L_N}{L_1}\right),
\end{align}
for $L_0 < L_N < L_1$, and $\epsilon_\text{eff} = 0$ otherwise. Here, $L_0 = 0.025$~m and $L_1 = 5$~m are the boundaries of the inner tracker of the FCC-hh as shown in Fig.~\ref{fig:detector}. Hence, we only consider the inner tracker in detecting displaced objects. The efficiency in Eq.~\eqref{lengthapproxi} peaks at $\epsilon_\text{eff}(L_N) \approx 10\%$ for $L_N \approx 0.5$~m, i.e., we use a reduction of the order of that used in LHC analyses \cite{ATLAS:2019kpx, ATL-PHYS-PUB-2019-013}. Such an order of magnitude reduction obviously worsens the FCC-hh reach but does not do so severely. As expected, the sensitivity is especially reduced in comparison with the $\epsilon_\text{eff} = 100\%$ case for smaller active-sterile mixing $|V_{\mu N}|$ as this corresponds to longer decay lengths.
	
As shown in Fig.~\ref{fig:deltaR}, the muons and jets in the final states tend to be very collimated, and we also illustrate the average $\overline{\Delta R}(\mu j)$ between the muon and closest jet in Fig.~\ref{fig:finite_eff}. For $\overline{\Delta R}(\mu j) \lesssim 0.4$, i.e., $m_N \lesssim 300$~GeV, different analyses strategies are beneficial due to the collimated muons and jets. This, in principle, applies to both the prompt and displaced $B-L$ $\mu jj$ signatures shown in Fig.~\ref{fig:finite_eff}. In the latter case, displaced fat jets can be considered which preserves sensitivity to small masses $m_N \gtrsim 10$~GeV \cite{Padhan:2022fak}. For the prompt signature, the interesting region with $m_N \gtrsim 100$~GeV and $|V_{\mu N}|\approx 10^{-7} - 10^{-6}$, motivated by the light neutrino masses, corresponds to fairly large $\overline{\Delta R}(\mu j) \gtrsim 0.1$.

%%%%%%%%%%%%%%%%%%%%%%%%%%%%%%%%%%%%%%%%%%%%%%%%
\section{Conclusion}
\label{sec:conclusion}
%%%%%%%%%%%%%%%%%%%%%%%%%%%%%%%%%%%%%%%%%%%%%%%%
The absence of an explanation of neutrino masses within the SM demands new physics. Experimental confirmation for such a mechanism will have a profound impact on understanding the fundamental laws of nature and will help build the next standard model. One such scenario is the seesaw mechanism in which heavy right handed neutrinos are predicted. These can be produced at the LHC via Drell-Yan processes mediated by a SM $W$ or a $Z$ boson. The reach of colliders such as LHC or FCC is however limited as the heavy neutrino production is suppressed by the active-sterile mixing which is expected to be small to generate the correct light neutrino masses. This observation necessitates exploring non-minimal heavy neutrino production as a means to probe neutrino mass generation.

We have here determined the sensitivity of the FCC-hh towards heavy neutrino production within the $B-L$ model, which is equipped with an additional $B-L$ gauge boson $Z'$. Using this gauge boson as a portal of heavy neutrino production, we explored the process $pp \to Z' \to NN$ and we have contrasted the resulting sensitivity against the assured but suppressed SM production channel $pp \to W \to l N$. We have concentrated on muon flavour focussing on two different neutrino decay modes, $N\to \mu\mu\nu$ and $N\to\mu jj$. The FCC-hh with 30~ab$^{-1}$ luminosity and 100~TeV center-of-mass energy can probe $B-L$ $Z'$ gauge bosons with masses of the order 50~TeV.
We use $m_{Z'} = 5$~TeV as benchmark with comparatively low values of the associated gauge coupling, namely $g_{B-L} = 10^{-2}$ and $0.1$. With such a $U(1)_{B-L}$ portal, the FCC-hh will be able to probe regions relevant for neutrino mass generation in the $\mu\mu\nu$ and the $\mu jj$ final state. Both prompt and displaced signals are relevant to cover a range of heavy neutrino mass $20~\text{GeV} \lesssim m_N \lesssim 2$~TeV with active-sterile mixing strength $|V_{\mu N}| \approx 10^{-7} - 10^{-5}$ motivated by light neutrino masses, $m_\nu \sim |V_{\mu N}|^2 m_N \sim 0.1$~eV. In the given scenario considered, active-sterile mixing strengths as low as $V_{\mu N} \approx 10^{-12}$ can be probed. Such searches for displaced heavy neutrinos have smaller background and thus may even be the first direct signal of an exotic $Z'$ resonance, potentially shedding light on the $R_K$ anomaly~\cite{Falkowski:2018dsl}.

The neutrino production processes mediated by the SM $W$ and $Z$, while assured and independent of the $U(1)_{B-L}$ extension, are suppressed by this active-sterile mixing. Only dedicated facilities with high fluxes and the potential to probe long lived particles such as SHiP and FCC-ee have the ability to probe the required small active-sterile mixing strength but only for limited ranges of neutrino masses, namely $m_N \approx 1-2$~GeV and $m_N \approx 50$~GeV, respectively. In the same channel, the FCC-hh will only be able to probe active-sterile mixing strengths $|V_{\mu N}| \approx 10^{-3} - 10^{-4}$, to high for a successful generic seesaw mechanism. This sensitivity is also comparable to the HL-LHC; the gain in terms of luminosity and cross section are compensated by the harder $p_T$ requirements.

While the mechanism of neutrino mass generation may be adapted to incorporate heavy neutrinos with large active-sterile mixing, such as in inverse seesaw scenarios, the vanilla seesaw remains an attractive and suggestive proposition. The corresponding Yukawa couplings between the left- and right-handed neutrinos may be small, $y_\nu \approx 10^{-6}$ for $m_N$ around the electroweak scale, but this is of the order of the Yukawa coupling of the electron. The minimal $B-L$ model has the appeal of incorporating the origin of light neutrino masses by breaking lepton number spontaneously. This is still possible near the electroweak scale and, as we have shown, it can lead to striking prompt and displaced signatures. Heavy neutrino production via a $Z'$ mediator has the potential to probe regions of parameter space relevant for neutrino mass generation in multiple final states. A refinement of our analysis with a full detector simulation should be able to demonstrate the reach more accurately, however, our first exploration demonstrates the potential of the FCC to target heavy neutrinos and the origin of light neutrinos.

\acknowledgments
%%%%%%%%%%%%%%%%%%%%%%%%%%%%%%%%%%%%%%%%%%%%%%%%%%%%%%%%%%%%%%%%%%%%%%%%%%%%%%%%%
WL is supported by the 2021 Jiangsu Shuangchuang (Mass Innovation and Entrepreneurship) Talent Program (JSSCBS20210213). SK is supported by Elise-Richter grant project number V592-N27 and FFD by a UK STFC consolidated grant (Reference ST/P00072X/1). The authors would like to thank F. Blekman, S. Pagan Griso and O. Fischer for several useful discussions.

\bibliographystyle{JHEP}
\bibliography{FCC_HNL.bib}

\providecommand{\href}[2]{#2}\begingroup\raggedright\begin{thebibliography}{10}

\bibitem{Davidson:1978pm}
A.~Davidson, \emph{{$B-L$ as the fourth color within an $\mathrm{SU}(2)_L
  \times \mathrm{U}(1)_R \times \mathrm{U}(1)$ model}},
  \href{http://dx.doi.org/10.1103/PhysRevD.20.776}{\emph{Phys. Rev. D} {\bf 20}
  (1979) 776}.

\bibitem{Mohapatra:1980qe}
R.~N. Mohapatra and R.~Marshak, \emph{{Local B-L Symmetry of Electroweak
  Interactions, Majorana Neutrinos and Neutron Oscillations}},
  \href{http://dx.doi.org/10.1103/PhysRevLett.44.1316}{\emph{Phys. Rev. Lett.}
  {\bf 44} (1980) 1316--1319}.

\bibitem{Chatrchyan:2012fla}
{\scshape CMS} collaboration, S.~Chatrchyan et~al., \emph{{Search for heavy
  Majorana Neutrinos in $\mu^{\pm}\mu^{\pm} +$ Jets and $e^{\pm}e^{\pm} +$ Jets
  Events in pp Collisions at $\sqrt{s} =$ 7 TeV}},
  \href{http://dx.doi.org/10.1016/j.physletb.2012.09.012}{\emph{Phys. Lett. B}
  {\bf 717} (2012) 109--128}, [\href{http://arxiv.org/abs/1207.6079}{{\tt
  1207.6079}}].

\bibitem{Aaij:2014aba}
{\scshape LHCb} collaboration, R.~Aaij et~al., \emph{{Search for Majorana
  neutrinos in $B^- \to \pi^+\mu^-\mu^-$ decays}},
  \href{http://dx.doi.org/10.1103/PhysRevLett.112.131802}{\emph{Phys. Rev.
  Lett.} {\bf 112} (2014) 131802}, [\href{http://arxiv.org/abs/1401.5361}{{\tt
  1401.5361}}].

\bibitem{Aad:2015xaa}
{\scshape ATLAS} collaboration, G.~Aad et~al., \emph{{Search for heavy Majorana
  neutrinos with the ATLAS detector in pp collisions at $ \sqrt{s}=8 $ TeV}},
  \href{http://dx.doi.org/10.1007/JHEP07(2015)162}{\emph{JHEP} {\bf 07} (2015)
  162}, [\href{http://arxiv.org/abs/1506.06020}{{\tt 1506.06020}}].

\bibitem{CMS:2015qur}
{\scshape CMS} collaboration, V.~Khachatryan et~al., \emph{{Search for heavy
  Majorana neutrinos in $\mu^\pm \mu^\pm+$ jets events in proton-proton
  collisions at $\sqrt{s}$ = 8 TeV}},
  \href{http://dx.doi.org/10.1016/j.physletb.2015.06.070}{\emph{Phys. Lett. B}
  {\bf 748} (2015) 144--166}, [\href{http://arxiv.org/abs/1501.05566}{{\tt
  1501.05566}}].

\bibitem{CMS:2016aro}
{\scshape CMS} collaboration, V.~Khachatryan et~al., \emph{{Search for heavy
  Majorana neutrinos in e$^{\pm}$e$^{\pm}$+ jets and e$^{\pm}$ $\mu^{\pm}$+
  jets events in proton-proton collisions at $ \sqrt{s}=8 $ TeV}},
  \href{http://dx.doi.org/10.1007/JHEP04(2016)169}{\emph{JHEP} {\bf 04} (2016)
  169}, [\href{http://arxiv.org/abs/1603.02248}{{\tt 1603.02248}}].

\bibitem{CortinaGil:2017mqf}
{\scshape NA62} collaboration, E.~Cortina~Gil et~al., \emph{{Search for heavy
  neutral lepton production in $K^+$ decays}},
  \href{http://dx.doi.org/10.1016/j.physletb.2018.01.031}{\emph{Phys. Lett. B}
  {\bf 778} (2018) 137--145}, [\href{http://arxiv.org/abs/1712.00297}{{\tt
  1712.00297}}].

\bibitem{Mermod:2017ceo}
{\scshape SHiP} collaboration, P.~Mermod, \emph{{Prospects of the SHiP and NA62
  experiments at CERN for hidden sector searches}},
  \href{http://dx.doi.org/10.22323/1.295.0139}{\emph{PoS} {\bf NuFact2017}
  (2017) 139}, [\href{http://arxiv.org/abs/1712.01768}{{\tt 1712.01768}}].

\bibitem{Izmaylov:2017lkv}
A.~Izmaylov and S.~Suvorov, \emph{{Search for heavy neutrinos in the ND280 near
  detector of the T2K experiment}},
  \href{http://dx.doi.org/10.1134/S1063779617060223}{\emph{Phys. Part. Nucl.}
  {\bf 48} (2017) 984--986}.

\bibitem{Sirunyan:2018mtv}
{\scshape CMS} collaboration, A.~M. Sirunyan et~al., \emph{{Search for heavy
  neutral leptons in events with three charged leptons in proton-proton
  collisions at $\sqrt{s} =$ 13 TeV}},
  \href{http://dx.doi.org/10.1103/PhysRevLett.120.221801}{\emph{Phys. Rev.
  Lett.} {\bf 120} (2018) 221801}, [\href{http://arxiv.org/abs/1802.02965}{{\tt
  1802.02965}}].

\bibitem{Helo:2010cw}
J.~C. Helo, S.~Kovalenko and I.~Schmidt, \emph{{Sterile neutrinos in lepton
  number and lepton flavor violating decays}},
  \href{http://dx.doi.org/10.1016/j.nuclphysb.2011.07.020}{\emph{Nucl. Phys. B}
  {\bf 853} (2011) 80--104}, [\href{http://arxiv.org/abs/1005.1607}{{\tt
  1005.1607}}].

\bibitem{Liventsev:2013zz}
{\scshape Belle} collaboration, D.~Liventsev et~al., \emph{{Search for heavy
  neutrinos at Belle}},
  \href{http://dx.doi.org/10.1103/PhysRevD.87.071102}{\emph{Phys. Rev. D} {\bf
  87} (2013) 071102}, [\href{http://arxiv.org/abs/1301.1105}{{\tt 1301.1105}}].

\bibitem{Abada:2013aba}
A.~Abada, A.~Teixeira, A.~Vicente and C.~Weiland, \emph{{Sterile neutrinos in
  leptonic and semileptonic decays}},
  \href{http://dx.doi.org/10.1007/JHEP02(2014)091}{\emph{JHEP} {\bf 02} (2014)
  091}, [\href{http://arxiv.org/abs/1311.2830}{{\tt 1311.2830}}].

\bibitem{Helo:2013esa}
J.~C. Helo, M.~Hirsch and S.~Kovalenko, \emph{{Heavy neutrino searches at the
  LHC with displaced vertices}},
  \href{http://dx.doi.org/10.1103/PhysRevD.89.073005}{\emph{Phys. Rev. D} {\bf
  89} (2014) 073005}, [\href{http://arxiv.org/abs/1312.2900}{{\tt 1312.2900}}].

\bibitem{Canetti:2014dka}
L.~Canetti, M.~Drewes and B.~Garbrecht, \emph{{Probing leptogenesis with
  GeV-scale sterile neutrinos at LHCb and Belle II}},
  \href{http://dx.doi.org/10.1103/PhysRevD.90.125005}{\emph{Phys. Rev. D} {\bf
  90} (2014) 125005}, [\href{http://arxiv.org/abs/1404.7114}{{\tt 1404.7114}}].

\bibitem{Gago:2015vma}
A.~M. Gago, P.~Hern\'andez, J.~Jones-P\'erez, M.~Losada and A.~Moreno
  Brice\~no, \emph{{Probing the Type I Seesaw Mechanism with Displaced Vertices
  at the LHC}},
  \href{http://dx.doi.org/10.1140/epjc/s10052-015-3693-1}{\emph{Eur. Phys. J.
  C} {\bf 75} (2015) 470}, [\href{http://arxiv.org/abs/1505.05880}{{\tt
  1505.05880}}].

\bibitem{Das:2015toa}
A.~Das and N.~Okada, \emph{{Improved bounds on the heavy neutrino productions
  at the LHC}}, \href{http://dx.doi.org/10.1103/PhysRevD.93.033003}{\emph{Phys.
  Rev. D} {\bf 93} (2016) 033003}, [\href{http://arxiv.org/abs/1510.04790}{{\tt
  1510.04790}}].

\bibitem{Banerjee:2015gca}
S.~Banerjee, P.~S.~B. Dev, A.~Ibarra, T.~Mandal and M.~Mitra, \emph{{Prospects
  of Heavy Neutrino Searches at Future Lepton Colliders}},
  \href{http://dx.doi.org/10.1103/PhysRevD.92.075002}{\emph{Phys. Rev. D} {\bf
  92} (2015) 075002}, [\href{http://arxiv.org/abs/1503.05491}{{\tt
  1503.05491}}].

\bibitem{Izaguirre:2015pga}
E.~Izaguirre and B.~Shuve, \emph{{Multilepton and Lepton Jet Probes of
  Sub-Weak-Scale Right-Handed Neutrinos}},
  \href{http://dx.doi.org/10.1103/PhysRevD.91.093010}{\emph{Phys. Rev. D} {\bf
  91} (2015) 093010}, [\href{http://arxiv.org/abs/1504.02470}{{\tt
  1504.02470}}].

\bibitem{Arganda:2015ija}
E.~Arganda, M.~Herrero, X.~Marcano and C.~Weiland, \emph{{Exotic
  \ensuremath{\mu}\ensuremath{\tau}jj events from heavy ISS neutrinos at the
  LHC}}, \href{http://dx.doi.org/10.1016/j.physletb.2015.11.013}{\emph{Phys.
  Lett. B} {\bf 752} (2016) 46--50},
  [\href{http://arxiv.org/abs/1508.05074}{{\tt 1508.05074}}].

\bibitem{Antusch:2015mia}
S.~Antusch and O.~Fischer, \emph{{Testing sterile neutrino extensions of the
  Standard Model at future lepton colliders}},
  \href{http://dx.doi.org/10.1007/JHEP05(2015)053}{\emph{JHEP} {\bf 05} (2015)
  053}, [\href{http://arxiv.org/abs/1502.05915}{{\tt 1502.05915}}].

\bibitem{Degrande:2016aje}
C.~Degrande, O.~Mattelaer, R.~Ruiz and J.~Turner, \emph{{Fully-Automated
  Precision Predictions for Heavy Neutrino Production Mechanisms at Hadron
  Colliders}}, \href{http://dx.doi.org/10.1103/PhysRevD.94.053002}{\emph{Phys.
  Rev. D} {\bf 94} (2016) 053002}, [\href{http://arxiv.org/abs/1602.06957}{{\tt
  1602.06957}}].

\bibitem{Antusch:2017pkq}
S.~Antusch, E.~Cazzato, M.~Drewes, O.~Fischer, B.~Garbrecht, D.~Gueter et~al.,
  \emph{{Probing Leptogenesis at Future Colliders}},
  \href{http://dx.doi.org/10.1007/JHEP09(2018)124}{\emph{JHEP} {\bf 09} (2018)
  124}, [\href{http://arxiv.org/abs/1710.03744}{{\tt 1710.03744}}].

\bibitem{Ruiz:2017yyf}
R.~Ruiz, M.~Spannowsky and P.~Waite, \emph{{Heavy neutrinos from gluon
  fusion}}, \href{http://dx.doi.org/10.1103/PhysRevD.96.055042}{\emph{Phys.
  Rev. D} {\bf 96} (2017) 055042}, [\href{http://arxiv.org/abs/1706.02298}{{\tt
  1706.02298}}].

\bibitem{Antusch:2017hhu}
S.~Antusch, E.~Cazzato and O.~Fischer, \emph{{Sterile neutrino searches via
  displaced vertices at LHCb}},
  \href{http://dx.doi.org/10.1016/j.physletb.2017.09.057}{\emph{Phys. Lett. B}
  {\bf 774} (2017) 114--118}, [\href{http://arxiv.org/abs/1706.05990}{{\tt
  1706.05990}}].

\bibitem{Dube:2017jgo}
S.~Dube, D.~Gadkari and A.~M. Thalapillil, \emph{{Lepton-Jets and Low-Mass
  Sterile Neutrinos at Hadron Colliders}},
  \href{http://dx.doi.org/10.1103/PhysRevD.96.055031}{\emph{Phys. Rev. D} {\bf
  96} (2017) 055031}, [\href{http://arxiv.org/abs/1707.00008}{{\tt
  1707.00008}}].

\bibitem{Cai:2017mow}
Y.~Cai, T.~Han, T.~Li and R.~Ruiz, \emph{{Lepton Number Violation: Seesaw
  Models and Their Collider Tests}},
  \href{http://dx.doi.org/10.3389/fphy.2018.00040}{\emph{Front. in Phys.} {\bf
  6} (2018) 40}, [\href{http://arxiv.org/abs/1711.02180}{{\tt 1711.02180}}].

\bibitem{Deppisch:2018eth}
F.~F. Deppisch, W.~Liu and M.~Mitra, \emph{{Long-lived Heavy Neutrinos from
  Higgs Decays}}, \href{http://dx.doi.org/10.1007/JHEP08(2018)181}{\emph{JHEP}
  {\bf 08} (2018) 181}, [\href{http://arxiv.org/abs/1804.04075}{{\tt
  1804.04075}}].

\bibitem{Abada:2018sfh}
A.~Abada, N.~Bernal, M.~Losada and X.~Marcano, \emph{{Inclusive Displaced
  Vertex Searches for Heavy Neutral Leptons at the LHC}},
  \href{http://dx.doi.org/10.1007/JHEP01(2019)093}{\emph{JHEP} {\bf 01} (2019)
  093}, [\href{http://arxiv.org/abs/1807.10024}{{\tt 1807.10024}}].

\bibitem{Cottin:2018nms}
G.~Cottin, J.~C. Helo and M.~Hirsch, \emph{{Displaced vertices as probes of
  sterile neutrino mixing at the LHC}},
  \href{http://dx.doi.org/10.1103/PhysRevD.98.035012}{\emph{Phys. Rev. D} {\bf
  98} (2018) 035012}, [\href{http://arxiv.org/abs/1806.05191}{{\tt
  1806.05191}}].

\bibitem{Drewes:2018gkc}
M.~Drewes, J.~Hajer, J.~Klaric and G.~Lanfranchi, \emph{{NA62 sensitivity to
  heavy neutral leptons in the low scale seesaw model}},
  \href{http://dx.doi.org/10.1007/JHEP07(2018)105}{\emph{JHEP} {\bf 07} (2018)
  105}, [\href{http://arxiv.org/abs/1801.04207}{{\tt 1801.04207}}].

\bibitem{Dib:2018iyr}
C.~O. Dib, C.~Kim, N.~A. Neill and X.-B. Yuan, \emph{{Search for sterile
  neutrinos decaying into pions at the LHC}},
  \href{http://dx.doi.org/10.1103/PhysRevD.97.035022}{\emph{Phys. Rev. D} {\bf
  97} (2018) 035022}, [\href{http://arxiv.org/abs/1801.03624}{{\tt
  1801.03624}}].

\bibitem{Boiarska:2019jcw}
I.~Boiarska, K.~Bondarenko, A.~Boyarsky, S.~Eijima, M.~Ovchynnikov,
  O.~Ruchayskiy et~al., \emph{{Probing baryon asymmetry of the Universe at LHC
  and SHiP}},  \href{http://arxiv.org/abs/1902.04535}{{\tt 1902.04535}}.

\bibitem{Cheung:2020buy}
K.~Cheung, Y.-L. Chung, H.~Ishida and C.-T. Lu, \emph{{Sensitivity reach on
  heavy neutral leptons and $\tau$-neutrino mixing $|U_{\tau N}|^2 $ at the
  HL-LHC}}, \href{http://dx.doi.org/10.1103/PhysRevD.102.075038}{\emph{Phys.
  Rev. D} {\bf 102} (2020) 075038},
  [\href{http://arxiv.org/abs/2004.11537}{{\tt 2004.11537}}].

\bibitem{Jones-Perez:2019plk}
J.~Jones-P\'erez, J.~Masias and J.~D. Ruiz-\'Alvarez, \emph{{Search for
  Long-Lived Heavy Neutrinos at the LHC with a VBF Trigger}},
  \href{http://dx.doi.org/10.1140/epjc/s10052-020-8188-z}{\emph{Eur. Phys. J.
  C} {\bf 80} (2020) 642}, [\href{http://arxiv.org/abs/1912.08206}{{\tt
  1912.08206}}].

\bibitem{Liu:2019ayx}
J.~Liu, Z.~Liu, L.-T. Wang and X.-P. Wang, \emph{{Seeking for sterile neutrinos
  with displaced leptons at the LHC}},
  \href{http://dx.doi.org/10.1007/JHEP07(2019)159}{\emph{JHEP} {\bf 07} (2019)
  159}, [\href{http://arxiv.org/abs/1904.01020}{{\tt 1904.01020}}].

\bibitem{Drewes:2019fou}
M.~Drewes and J.~Hajer, \emph{{Heavy Neutrinos in displaced vertex searches at
  the LHC and HL-LHC}},
  \href{http://dx.doi.org/10.1007/JHEP02(2020)070}{\emph{JHEP} {\bf 02} (2020)
  070}, [\href{http://arxiv.org/abs/1903.06100}{{\tt 1903.06100}}].

\bibitem{Alekhin:2015byh}
S.~Alekhin et~al., \emph{{A facility to Search for Hidden Particles at the CERN
  SPS: the SHiP physics case}},
  \href{http://dx.doi.org/10.1088/0034-4885/79/12/124201}{\emph{Rept. Prog.
  Phys.} {\bf 79} (2016) 124201}, [\href{http://arxiv.org/abs/1504.04855}{{\tt
  1504.04855}}].

\bibitem{SHiP:2018xqw}
{\scshape SHiP} collaboration, C.~Ahdida et~al., \emph{{Sensitivity of the SHiP
  experiment to Heavy Neutral Leptons}},
  \href{http://dx.doi.org/10.1007/JHEP04(2019)077}{\emph{JHEP} {\bf 04} (2019)
  077}, [\href{http://arxiv.org/abs/1811.00930}{{\tt 1811.00930}}].

\bibitem{Das:2018usr}
A.~Das, S.~Jana, S.~Mandal and S.~Nandi, \emph{{Probing right handed neutrinos
  at the LHeC and lepton colliders using fat jet signatures}},
  \href{http://dx.doi.org/10.1103/PhysRevD.99.055030}{\emph{Phys. Rev. D} {\bf
  99} (2019) 055030}, [\href{http://arxiv.org/abs/1811.04291}{{\tt
  1811.04291}}].

\bibitem{Antusch:2016ejd}
S.~Antusch, E.~Cazzato and O.~Fischer, \emph{{Sterile neutrino searches at
  future $e^-e^+$, $pp$, and $e^-p$ colliders}},
  \href{http://dx.doi.org/10.1142/S0217751X17500786}{\emph{Int. J. Mod. Phys.
  A} {\bf 32} (2017) 1750078}, [\href{http://arxiv.org/abs/1612.02728}{{\tt
  1612.02728}}].

\bibitem{Pascoli:2018heg}
S.~Pascoli, R.~Ruiz and C.~Weiland, \emph{{Heavy neutrinos with dynamic jet
  vetoes: multilepton searches at $ \sqrt{s}=14 $ , 27, and 100 TeV}},
  \href{http://dx.doi.org/10.1007/JHEP06(2019)049}{\emph{JHEP} {\bf 06} (2019)
  049}, [\href{http://arxiv.org/abs/1812.08750}{{\tt 1812.08750}}].

\bibitem{Chiang:2019ajm}
C.-W. Chiang, G.~Cottin, A.~Das and S.~Mandal, \emph{{Displaced heavy neutrinos
  from $Z^\prime$ decays at the LHC}},
  \href{http://dx.doi.org/10.1007/JHEP12(2019)070}{\emph{JHEP} {\bf 12} (2019)
  070}, [\href{http://arxiv.org/abs/1908.09838}{{\tt 1908.09838}}].

\bibitem{Deppisch:2013cya}
F.~F. Deppisch, N.~Desai and J.~W.~F. Valle, \emph{{Is charged lepton flavor
  violation a high energy phenomenon?}},
  \href{http://dx.doi.org/10.1103/PhysRevD.89.051302}{\emph{Phys. Rev. D} {\bf
  89} (2014) 051302}, [\href{http://arxiv.org/abs/1308.6789}{{\tt 1308.6789}}].

\bibitem{Batell:2016zod}
B.~Batell, M.~Pospelov and B.~Shuve, \emph{{Shedding Light on Neutrino Masses
  with Dark Forces}},
  \href{http://dx.doi.org/10.1007/JHEP08(2016)052}{\emph{JHEP} {\bf 08} (2016)
  052}, [\href{http://arxiv.org/abs/1604.06099}{{\tt 1604.06099}}].

\bibitem{Deppisch:2019kvs}
F.~Deppisch, S.~Kulkarni and W.~Liu, \emph{{Heavy neutrino production via
  $Z^{\prime}$ at the lifetime frontier}},
  \href{http://dx.doi.org/10.1103/PhysRevD.100.035005}{\emph{Phys. Rev. D} {\bf
  100} (2019) 035005}, [\href{http://arxiv.org/abs/1905.11889}{{\tt
  1905.11889}}].

\bibitem{Bhattacherjee:2021rml}
B.~Bhattacherjee, S.~Matsumoto and R.~Sengupta, \emph{{Long-Lived Light
  Mediators from Higgs boson Decay at HL-LHC, FCC-hh and a Proposal of
  Dedicated LLP Detectors for FCC-hh}},
  \href{http://arxiv.org/abs/2111.02437}{{\tt 2111.02437}}.

\bibitem{Accomando:2017qcs}
E.~Accomando, L.~Delle~Rose, S.~Moretti, E.~Olaiya and C.~H.
  Shepherd-Themistocleous, \emph{{Extra Higgs boson and Z$^{\prime}$ as portals
  to signatures of heavy neutrinos at the LHC}},
  \href{http://dx.doi.org/10.1007/JHEP02(2018)109}{\emph{JHEP} {\bf 02} (2018)
  109}, [\href{http://arxiv.org/abs/1708.03650}{{\tt 1708.03650}}].

\bibitem{Das:2019fee}
A.~Das, P.~B. Dev and N.~Okada, \emph{{Long-lived TeV-scale right-handed
  neutrino production at the LHC in gauged $U(1)_X$ model}},
  \href{http://dx.doi.org/10.1016/j.physletb.2019.135052}{\emph{Phys. Lett. B}
  {\bf 799} (2019) 135052}, [\href{http://arxiv.org/abs/1906.04132}{{\tt
  1906.04132}}].

\bibitem{Cheung:2021utb}
K.~Cheung, K.~Wang and Z.~S. Wang, \emph{{Time-delayed electrons from neutral
  currents at the LHC}},
  \href{http://dx.doi.org/10.1007/JHEP09(2021)026}{\emph{JHEP} {\bf 09} (2021)
  026}, [\href{http://arxiv.org/abs/2107.03203}{{\tt 2107.03203}}].

\bibitem{FileviezPerez:2020cgn}
P.~Fileviez~P\'erez and A.~D. Plascencia, \emph{{Probing the Nature of
  Neutrinos with a New Force}},
  \href{http://dx.doi.org/10.1103/PhysRevD.102.015010}{\emph{Phys. Rev. D} {\bf
  102} (2020) 015010}, [\href{http://arxiv.org/abs/2005.04235}{{\tt
  2005.04235}}].

\bibitem{Accomando:2016rpc}
E.~Accomando, L.~Delle~Rose, S.~Moretti, E.~Olaiya and C.~H.
  Shepherd-Themistocleous, \emph{{Novel SM-like Higgs decay into displaced
  heavy neutrino pairs in U(1)' models}},
  \href{http://dx.doi.org/10.1007/JHEP04(2017)081}{\emph{JHEP} {\bf 04} (2017)
  081}, [\href{http://arxiv.org/abs/1612.05977}{{\tt 1612.05977}}].

\bibitem{Han:2021pun}
C.~Han, T.~Li and C.-Y. Yao, \emph{{Searching for heavy neutrino in terms of
  tau lepton at future hadron collider}},
  \href{http://dx.doi.org/10.1103/PhysRevD.104.015036}{\emph{Phys. Rev. D} {\bf
  104} (2021) 015036}, [\href{http://arxiv.org/abs/2103.03548}{{\tt
  2103.03548}}].

\bibitem{Deppisch:2010fr}
F.~F. Deppisch and A.~Pilaftsis, \emph{{Lepton Flavour Violation and theta(13)
  in Minimal Resonant Leptogenesis}},
  \href{http://dx.doi.org/10.1103/PhysRevD.83.076007}{\emph{Phys. Rev. D} {\bf
  83} (2011) 076007}, [\href{http://arxiv.org/abs/1012.1834}{{\tt 1012.1834}}].

\bibitem{Atre:2009rg}
A.~Atre, T.~Han, S.~Pascoli and B.~Zhang, \emph{{The Search for Heavy Majorana
  Neutrinos}},
  \href{http://dx.doi.org/10.1088/1126-6708/2009/05/030}{\emph{JHEP} {\bf 05}
  (2009) 030}, [\href{http://arxiv.org/abs/0901.3589}{{\tt 0901.3589}}].

\bibitem{FCC:2018vvp}
{\scshape FCC} collaboration, A.~Abada et~al., \emph{{FCC-hh: The Hadron
  Collider}: {Future Circular Collider Conceptual Design Report Volume 3}},
  \href{http://dx.doi.org/10.1140/epjst/e2019-900087-0}{\emph{Eur. Phys. J. ST}
  {\bf 228} (2019) 755--1107}.

\bibitem{private}
F.~Blekman, \emph{{private communication}}, .

\bibitem{Degrande:2011ua}
C.~Degrande, C.~Duhr, B.~Fuks, D.~Grellscheid, O.~Mattelaer and T.~Reiter,
  \emph{{UFO - The Universal FeynRules Output}},
  \href{http://dx.doi.org/10.1016/j.cpc.2012.01.022}{\emph{Comput. Phys.
  Commun.} {\bf 183} (2012) 1201--1214},
  [\href{http://arxiv.org/abs/1108.2040}{{\tt 1108.2040}}].

\bibitem{Alwall:2014hca}
J.~Alwall, R.~Frederix, S.~Frixione, V.~Hirschi, F.~Maltoni, O.~Mattelaer
  et~al., \emph{{The automated computation of tree-level and next-to-leading
  order differential cross sections, and their matching to parton shower
  simulations}}, \href{http://dx.doi.org/10.1007/JHEP07(2014)079}{\emph{JHEP}
  {\bf 07} (2014) 079}, [\href{http://arxiv.org/abs/1405.0301}{{\tt
  1405.0301}}].

\bibitem{Alloul:2013bka}
A.~Alloul, N.~D. Christensen, C.~Degrande, C.~Duhr and B.~Fuks,
  \emph{{FeynRules 2.0 - A complete toolbox for tree-level phenomenology}},
  \href{http://dx.doi.org/10.1016/j.cpc.2014.04.012}{\emph{Comput. Phys.
  Commun.} {\bf 185} (2014) 2250--2300},
  [\href{http://arxiv.org/abs/1310.1921}{{\tt 1310.1921}}].

\bibitem{Christensen:2008py}
N.~D. Christensen and C.~Duhr, \emph{{FeynRules - Feynman rules made easy}},
  \href{http://dx.doi.org/10.1016/j.cpc.2009.02.018}{\emph{Comput. Phys.
  Commun.} {\bf 180} (2009) 1614--1641},
  [\href{http://arxiv.org/abs/0806.4194}{{\tt 0806.4194}}].

\bibitem{FeynrulesDatabase}
``Feynrulesdatabase.'' \url{https://feynrules.irmp.ucl.ac.be/wiki/B-L-SM}.

\bibitem{Sjostrand:2014zea}
T.~Sj$\ddot{\text{o}}$strand, S.~Ask, J.~R. Christiansen, R.~Corke, N.~Desai,
  P.~Ilten et~al., \emph{{An Introduction to PYTHIA 8.2}},
  \href{http://dx.doi.org/10.1016/j.cpc.2015.01.024}{\emph{Comput. Phys.
  Commun.} {\bf 191} (2015) 159--177},
  [\href{http://arxiv.org/abs/1410.3012}{{\tt 1410.3012}}].

\bibitem{Cacciari:2011ma}
M.~Cacciari, G.~P. Salam and G.~Soyez, \emph{{FastJet User Manual}},
  \href{http://dx.doi.org/10.1140/epjc/s10052-012-1896-2}{\emph{Eur. Phys. J.
  C} {\bf 72} (2012) 1896}, [\href{http://arxiv.org/abs/1111.6097}{{\tt
  1111.6097}}].

\bibitem{CMS:2019tbu}
{\scshape CMS} collaboration, \emph{{Search for a narrow resonance in high-mass
  dilepton final states in proton-proton collisions using
  140$~\mathrm{fb}^{-1}$ of data at $\sqrt{s}=13~\mathrm{TeV}$}}, .

\bibitem{CMS:2021ctt}
{\scshape CMS} collaboration, A.~M. Sirunyan et~al., \emph{{Search for resonant
  and nonresonant new phenomena in high-mass dilepton final states at $
  \sqrt{s} $ = 13 TeV}},
  \href{http://dx.doi.org/10.1007/JHEP07(2021)208}{\emph{JHEP} {\bf 07} (2021)
  208}, [\href{http://arxiv.org/abs/2103.02708}{{\tt 2103.02708}}].

\bibitem{Helsens:2019bfw}
C.~Helsens, D.~Jamin, M.~L. Mangano, T.~G. Rizzo and M.~Selvaggi, \emph{{Heavy
  resonances at energy-frontier hadron colliders}},
  \href{http://dx.doi.org/10.1140/epjc/s10052-019-7062-3}{\emph{Eur. Phys. J.
  C} {\bf 79} (2019) 569}, [\href{http://arxiv.org/abs/1902.11217}{{\tt
  1902.11217}}].

\bibitem{Deppisch:2019ldi}
F.~F. Deppisch, S.~Kulkarni and W.~Liu, \emph{{Searching for a light
  $Z^{\prime}$ through Higgs production at the LHC}},
  \href{http://dx.doi.org/10.1103/PhysRevD.100.115023}{\emph{Phys. Rev. D} {\bf
  100} (2019) 115023}, [\href{http://arxiv.org/abs/1908.11741}{{\tt
  1908.11741}}].

\bibitem{Amrith:2018yfb}
S.~Amrith, J.~M. Butterworth, F.~F. Deppisch, W.~Liu, A.~Varma and D.~Yallup,
  \emph{{LHC Constraints on a $B-L$ Gauge Model using Contur}},
  \href{http://dx.doi.org/10.1007/JHEP05(2019)154}{\emph{JHEP} {\bf 05} (2019)
  154}, [\href{http://arxiv.org/abs/1811.11452}{{\tt 1811.11452}}].

\bibitem{Aad:2019fac}
{\scshape ATLAS} collaboration, G.~Aad et~al., \emph{{Search for high-mass
  dilepton resonances using 139 fb$^{-1}$ of $pp$ collision data collected at
  $\sqrt{s}=$13 TeV with the ATLAS detector}},
  \href{http://dx.doi.org/10.1016/j.physletb.2019.07.016}{\emph{Phys. Lett. B}
  {\bf 796} (2019) 68--87}, [\href{http://arxiv.org/abs/1903.06248}{{\tt
  1903.06248}}].

\bibitem{ATLAS:2019jvq}
{\scshape ATLAS} collaboration, M.~Aaboud et~al., \emph{{Electron
  reconstruction and identification in the ATLAS experiment using the 2015 and
  2016 LHC proton-proton collision data at $\sqrt{s}$ = 13 TeV}},
  \href{http://dx.doi.org/10.1140/epjc/s10052-019-7140-6}{\emph{Eur. Phys. J.
  C} {\bf 79} (2019) 639}, [\href{http://arxiv.org/abs/1902.04655}{{\tt
  1902.04655}}].

\bibitem{ATLAS:2019kpx}
{\scshape ATLAS} collaboration, G.~Aad et~al., \emph{{Search for heavy neutral
  leptons in decays of $W$ bosons produced in 13 TeV $pp$ collisions using
  prompt and displaced signatures with the ATLAS detector}},
  \href{http://dx.doi.org/10.1007/JHEP10(2019)265}{\emph{JHEP} {\bf 10} (2019)
  265}, [\href{http://arxiv.org/abs/1905.09787}{{\tt 1905.09787}}].

\bibitem{LHCb:2020akw}
{\scshape LHCb} collaboration, R.~Aaij et~al., \emph{{Search for long-lived
  particles decaying to $e^\pm \mu^\mp \nu$}},
  \href{http://dx.doi.org/10.1140/epjc/s10052-021-08994-0}{\emph{Eur. Phys. J.
  C} {\bf 81} (2021) 261}, [\href{http://arxiv.org/abs/2012.02696}{{\tt
  2012.02696}}].

\bibitem{CMS:2022fut}
{\scshape CMS} collaboration, A.~Tumasyan et~al., \emph{{Search for long-lived
  heavy neutral leptons with displaced vertices in proton-proton collisions at
  $\sqrt{s}$ =13 TeV}},  \href{http://arxiv.org/abs/2201.05578}{{\tt
  2201.05578}}.

\bibitem{Bolton:2019pcu}
P.~D. Bolton, F.~F. Deppisch and P.~S. Bhupal~Dev, \emph{{Neutrinoless double
  beta decay versus other probes of heavy sterile neutrinos}},
  \href{http://dx.doi.org/10.1007/JHEP03(2020)170}{\emph{JHEP} {\bf 03} (2020)
  170}, [\href{http://arxiv.org/abs/1912.03058}{{\tt 1912.03058}}].

\bibitem{Blondel:2014bra}
{\scshape FCC-ee study Team} collaboration, A.~Blondel, E.~Graverini, N.~Serra
  and M.~Shaposhnikov, \emph{{Search for Heavy Right Handed Neutrinos at the
  FCC-ee}},
  \href{http://dx.doi.org/10.1016/j.nuclphysbps.2015.09.304}{\emph{Nucl. Part.
  Phys. Proc.} {\bf 273-275} (2016) 1883--1890},
  [\href{http://arxiv.org/abs/1411.5230}{{\tt 1411.5230}}].

\bibitem{Aaboud:2017buh}
{\scshape ATLAS} collaboration, M.~Aaboud et~al., \emph{{Search for new
  high-mass phenomena in the dilepton final state using 36 fb$^{-1}$ of
  proton-proton collision data at $ \sqrt{s}=13 $ TeV with the ATLAS
  detector}}, \href{http://dx.doi.org/10.1007/JHEP10(2017)182}{\emph{JHEP} {\bf
  10} (2017) 182}, [\href{http://arxiv.org/abs/1707.02424}{{\tt 1707.02424}}].

\bibitem{Aad:2019kiz}
{\scshape ATLAS} collaboration, G.~Aad et~al., \emph{{Search for heavy neutral
  leptons in decays of $W$ bosons produced in 13 TeV $pp$ collisions using
  prompt and displaced signatures with the ATLAS detector}},
  \href{http://dx.doi.org/10.1007/JHEP10(2019)265}{\emph{JHEP} {\bf 10} (2019)
  265}, [\href{http://arxiv.org/abs/1905.09787}{{\tt 1905.09787}}].

\bibitem{ATL-PHYS-PUB-2019-013}
{\scshape ATLAS Collaboration} collaboration, \emph{{Performance of vertex
  reconstruction algorithms for detection of new long-lived particle decays
  within the ATLAS inner detector}},  tech. rep., CERN, Geneva.

\bibitem{Padhan:2022fak}
R.~Padhan, M.~Mitra, S.~Kulkarni and F.~F. Deppisch, \emph{{Displaced fat-jets
  and tracks to probe boosted right-handed neutrinos in the $U(1)_{B-L}$
  model}},  in \emph{{2022 Snowmass Summer Study}}, 3, 2022.
\newblock \href{http://arxiv.org/abs/2203.06114}{{\tt 2203.06114}}.

\bibitem{Falkowski:2018dsl}
A.~Falkowski, S.~F. King, E.~Perdomo and M.~Pierre, \emph{{Flavourful $Z'$
  portal for vector-like neutrino Dark Matter and $R_{K^{(*)}}$}},
  \href{http://dx.doi.org/10.1007/JHEP08(2018)061}{\emph{JHEP} {\bf 08} (2018)
  061}, [\href{http://arxiv.org/abs/1803.04430}{{\tt 1803.04430}}].

\end{thebibliography}\endgroup
\end{document}